\pdfoutput=1
\documentclass[a4paper, superscriptaddress, amsfonts, amssymb, amsmath, reprint, showkeys, nofootinbib, twoside]{revtex4-1}

\usepackage{graphicx}% Include figure files
\usepackage{dcolumn}% Align table columns on decimal point
\usepackage{bm}% bold math
\usepackage{enumitem}
\usepackage{graphicx}
\usepackage{amsthm}
\usepackage{subfigure}
\usepackage{multirow}
\usepackage[pdftex, pdftitle={Article}, pdfauthor={Author}]{hyperref} % For hyperlinks in the PDF
\usepackage{xcolor}
\usepackage{refcount}

\usepackage{amsmath}
\newcommand{\bs}{\boldsymbol}
\newcommand{\Hquad}{\hspace{0.5em}} 

\newcommand{\mc}{\mathcal}
\newcommand*\diff{\mathop{}\!\mathrm{d}}
\newcommand{\overbar}[1]{\mkern 1.5mu\overline{\mkern-1.5mu#1\mkern-1.5mu}\mkern 1.5mu}

\newcommand{\addrone}{Australian Research Council Centre of Excellence for Gravitational Wave Discovery (OzGrav) }
\newcommand{\addrtwo}{Department of physics, University of Western Australia, Crawley WA 6009, Australia}
\newcommand{\addrthr}{Gravitational Wave Data Centre, Swinburne University, Hawthorn VIC 3122, Australia}
\newcommand{\addrfour}{Department of Computer Science and Technology, Tsinghua University, Beijing, China}
\newcommand{\addrfive}{Theoretical Astrophysics 350-17, California Institute of Technology, Pasadena, CA 91125, USA}
\newcommand{\addrsix}{School of Physics and Astronomy, Monash University, Clayton, Victoria 3800, Australia}
\newcommand{\addrseven}{Department of Computer Science, New Jersey Institute of Technology, New Jersey 07102, USA}
\newcommand{\addreight}{Department of Physics, Indian Institute of Technology Gandhinagar, Gujarat 382355, India}
\newcommand{\addrnine}{Department of Physics, The Chinese University of Hong Kong, Shatin, N.T., Hong Kong}
\newcommand{\addrten}{Centre for Gravitational Waves, Institute for Theoretical Physics, KU Leuven, Celestijnenlaan 200D, 3001 Leuven, Belgium}
\newcommand{\addreleven}{Korea Astronomy and Space Science Institute, 776 Daedeokdae-ro, Yuseong-gu, Daejeon 34055, Republic of Korea}
\newcommand{\addrtwelve}{School of Physics, Australian National University, Acton Australian Capital Territory 2601, Australia}
\newcommand{\addrthirteen}{Centre for Astrophysics and Supercomputing, Swinburne University, Hawthorn Victoria 3122, Australia}

\begin{document}
\title{The SPIIR online coherent pipeline to search for gravitational waves from compact binary coalescences}

\author{Qi Chu}
\email{qi.chu@uwa.edu.au}
\affiliation{\addrone} 
\affiliation{\addrtwo}

\author{ Manoj Kovalam}
\affiliation{\addrone} 
\affiliation{\addrtwo}

\author{Linqing Wen}
\email{linqing.wen@uwa.edu.au}
\affiliation{\addrone} 
\affiliation{\addrtwo}

\author{Teresa Slaven-Blair}
\affiliation{\addrone} 
\affiliation{\addrtwo}

\author{Joel Bosveld}
\affiliation{\addrtwo}

\author{Yanbei Chen}
\affiliation{\addrfive}

\author{Patrick Clearwater}
\affiliation{\addrone}
\affiliation{\addrthr}

\author{Alex Codoreanu}
\affiliation{\addrone}
\affiliation{\addrthr}

\author{Zhihui Du}
\affiliation{\addrseven}

\author{Xiangyu Guo}
\affiliation{\addrfour}

\author{Xiaoyang Guo}
\affiliation{\addrfour}

\author{Kyungmin Kim}
\affiliation{\addreleven}

\author{Tjonnie G. F. Li}
\affiliation{\addrnine}
\affiliation{\addrten}

\author{Victor Oloworaran}
\affiliation{\addrtwo}

\author{Fiona Panther}
\affiliation{\addrone}
\affiliation{\addrtwo}

\author{Jade Powell}
\affiliation{\addrone}
\affiliation{\addrthirteen}

\author{Anand S. Sengupta}
\affiliation{\addreight}

\author{Karl Wette}
\affiliation{\addrone}
\affiliation{\addrtwelve}

\author{Xingjiang Zhu}
\affiliation{\addrone}
\affiliation{\addrsix}

\date{\today}
\begin{abstract}
This paper presents the SPIIR pipeline used for public alerts during the third advanced LIGO and Virgo observation run (O3 run). The SPIIR pipeline uses infinite impulse response (IIR) filters to perform extremely low-latency matched filtering and this process is further accelerated with graphics processing units (GPUs). It is the first online pipeline to select candidates from multiple detectors using a coherent statistic based on the maximum network likelihood ratio statistic principle. Here we simplify the derivation of this statistic using the singular-value-decomposition (SVD) technique and show that single-detector signal-to-noise ratios from matched filtering can be directly used to construct the statistic for each sky direction. Coherent searches are in general more computationally challenging than coincidence searches due to extra search over sky direction parameters. The search over sky directions follows an embarrassing parallelization paradigm and has been accelerated using GPUs. The detection performance is reported using a segment of public data from LIGO-Virgo's second observation run. We demonstrate that the median latency of the SPIIR pipeline is less than 9 seconds, and present an achievable roadmap to reduce the latency to less than 5 seconds. During the O3 online run, SPIIR registered triggers associated with 38 of the 56 non-retracted public alerts. The extreme low-latency nature makes it a competitive choice for joint time-domain observations, and offers the tantalizing possibility of making public alerts prior to the merger phase of binary coalescence systems involving at least one neutron star.

\end{abstract}
\maketitle

\section{Introduction}

Gravitational wave (GW) astronomy has been advancing rapidly since the first operation (referred to as O1) of the two Advanced Laser Interferometer Gravitational-wave Observatory (aLIGO) in 2015~\cite{aLIGO_2015}.  The advanced Virgo detector~\cite{AdV_2015} joined the LIGO detectors from the second observation run (O2). The third and latest run (O3) of aLIGO and Virgo lasted 11 months and finished in March 2020~\cite{LIGO_O3, Virgo_O3}. There have been over a dozen detections of compact binary coalescences (CBCs) by the LIGO collaboration in the O1 and O2 run~\cite{gwtc1, Nitz2020, barakO2}. The first six months of O3 has seen three times more detections than O1 and O2 combined~\cite{gwtc2}. With new advanced detectors, the KAGRA detector in operation since 2020~\cite{KAGRA_2020} and LIGO-India envisioned to be operational in the next decade~\cite{ligo_india}, it is expected the GW astronomy will see regular frequent detections and possible new breakthroughs.

%Among the significant detections, there is by far one detection of GW from merging binary neutron stars (GW170817~\cite{gw170817}) which had multiple coincident observations across electromagnetic (EM) spectrum~\cite{gw170817_mma}.
%It has seen a number of breakthroughs, including some highlights: the first direct detection of GWs (the binary black hole merger GW150914~\cite{gw150914}), the first detection of GW from merging binary neutron stars (GW170817~\cite{gw170817}) which had multiple coincident observations across electromagnetic (EM) spectrum~\cite{gw170817_mma}, and the first GW from merging intermediate-mass black hole (GW190521~\cite{gw190521}). 

A high priority for observing runs is to use GW detections as a trigger for electromagnetic (EM) follow-up observations. A successful example is the first detection of GW from merging binary neutron stars (GW170817~\cite{gw170817}) with multiple coincident observations across electromagnetic spectrum~\cite{gw170817_mma}. Improved detector sensitivity will expose more binary merger signals, as well as enabling detections ahead of the final coalescence phase, known as the early warning detections. To facilitate the real-time and early-warning detections, a public alert infrastructure\footnote{\url{https://emfollow.docs.ligo.org/userguide/}} was established by LIGO-Virgo collaboration before O3 to enable online data streaming, pipeline detection and prompt real-time alert publication.

%We briefly review the O3 public alert infrastructure. At the first stage, data were streamed live from the detector sites. The data stream included readouts from the interferometer itself (i.e. the main strain channel), as well as auxiliary channels from instrument monitors and noise monitors. The data in the strain channel was then calibrated and the key auxiliary channels are aggregated to generate `data quality' (DQ) channels to inform the online pipelines. This stage currently accrued ten seconds from the two LIGO detectors. This latency is expected to be reduced for the next observing run O4.

%Following that, the online data stream was distributed to computing clusters where online pipelines were operating. Each pipeline analyzed the data and produced candidates of either modeled events or unmodeled burst events that are subsequently sent to the gravitational-wave candidate event database (GraceDB\footnote{\url{https://gracedb.ligo.org/}}).  If a candidate was significant enough, the automated trigger orchestra program \texttt{GWCelery}\footnote{\url{https://gwcelery.readthedocs.io/}} would then collect triggers from all pipelines and selected the one with the maximum signal-to-noise ratio (SNR) (named as ``superevents"). This candidate would then be published to the public page of GraceDB and produce a preliminary GCN notice\footnote{\url{https://gcn.gsfc.nasa.gov/}}. Automated sky localization ~\cite{bayestar,veitch15} and source type estimation~\cite{pastro_method} would be triggered and released, or on the contrary a retraction would be issued based on low-latency noise investigations, bringing the end to an online detection.

Five detection pipelines have been used in this infrastructure: one to search for unmodeled signals, the cWB pipeline~\cite{klimenko2016}; four to search for the modeled compact binary coalescence (CBC) signals, including GstLAL \cite{gstlal_paper,Sachdev:2019vvd,Hanna:2019ezx}, MBTA \cite{MBTA_pipeline}, PyCBCLive~\cite{pycbc2017,Nitz:2018rgo,DalCanton:2020vpm}, and the SPIIR pipeline which is the focus of this paper. Previous work on the SPIIR pipeline development can be found in~\cite{jing,spiir_shaun,david_thesis,liuyuan,xiangyu2018,xiaoyang2018} and the SPIIR pipeline was used instrumentally during the O1 and O2 runs.

The SPIIR pipeline is distinguished from other CBC search pipelines in several aspects. It adopts the summed parallel infinite impulse response (SPIIR) method for matched filtering~\cite{jing,spiir_shaun,david_thesis}. This method is expected to be more efficient computationally than the traditional Fourier method when a filtering delay of less than $10\mathrm{s}$ is intended~\cite{jing}. It is straightforward to parallelize this algorithm using Graphics Processing Units (GPUs), a popular and cost-effective parallel computing platform~\cite{liuyuan,xiangyu2018}.

The other main difference is that the SPIIR pipeline selects candidates based on the maximum network likelihood ratio principle which is referred as the coherent method. Coherent methods have been developed for periodic GW searches~\cite{jaranowski98}, inspiral searches in the band of the proposed space-based interferometric detector, LISA~\cite{rogan2004,krolak2004,neil2006}, and for GRB-triggered CBC searches~\cite{pygrb,harry2011}. It was proposed for CBC searches~\cite{pai01,bose2011,macleod2016} but not widely used due to computational challenges of searching through additional parameters of source sky directions. A recent work~\cite{coh_PSO2020} is proposed to reduce the computational cost of parameter search using particle swarm optimization. In this paper, we express the coherent method for CBC signals using singular value decomposition (SVD). SVD and its variation principal component analysis have been applied to many areas of GW research including waveform decomposition~\cite{kipp2011,Heng_2009} and parameter estimation~\cite{rover2009}. It has been proposed for general formalization of GW data analysis with a detector network~\cite{wen2008}. The SVD derivation here is an extension of~\cite{wen2008} and simplifies the expression of the coherent statistic for CBC searches. It shows that output from matched filtering, i.e. the signal-to-noise ratio (SNR) time series, can be directly used and only the two parameters of sky directions need to be searched over. The search on each sky direction is independent and thus has been distributed to parallel GPU threads for acceleration.

This paper is organized as follows: Sec.~\ref{sec:spiir_overview} gives a detailed explanation of the SPIIR pipeline.
Sec.~\ref{sec:results} reports the performance of the SPIIR pipeline using a segment of the public O2 data, including a break-down analysis of contributions to the pipeline latency. It also reports the results of the pipeline on O3 public alerts. Sec.~\ref{sec:conclusion} gives conclusion and future perspectives of the pipeline. 

\section{Pipeline description}
\label{sec:spiir_overview}

\begin{figure}
\includegraphics[width=\columnwidth]{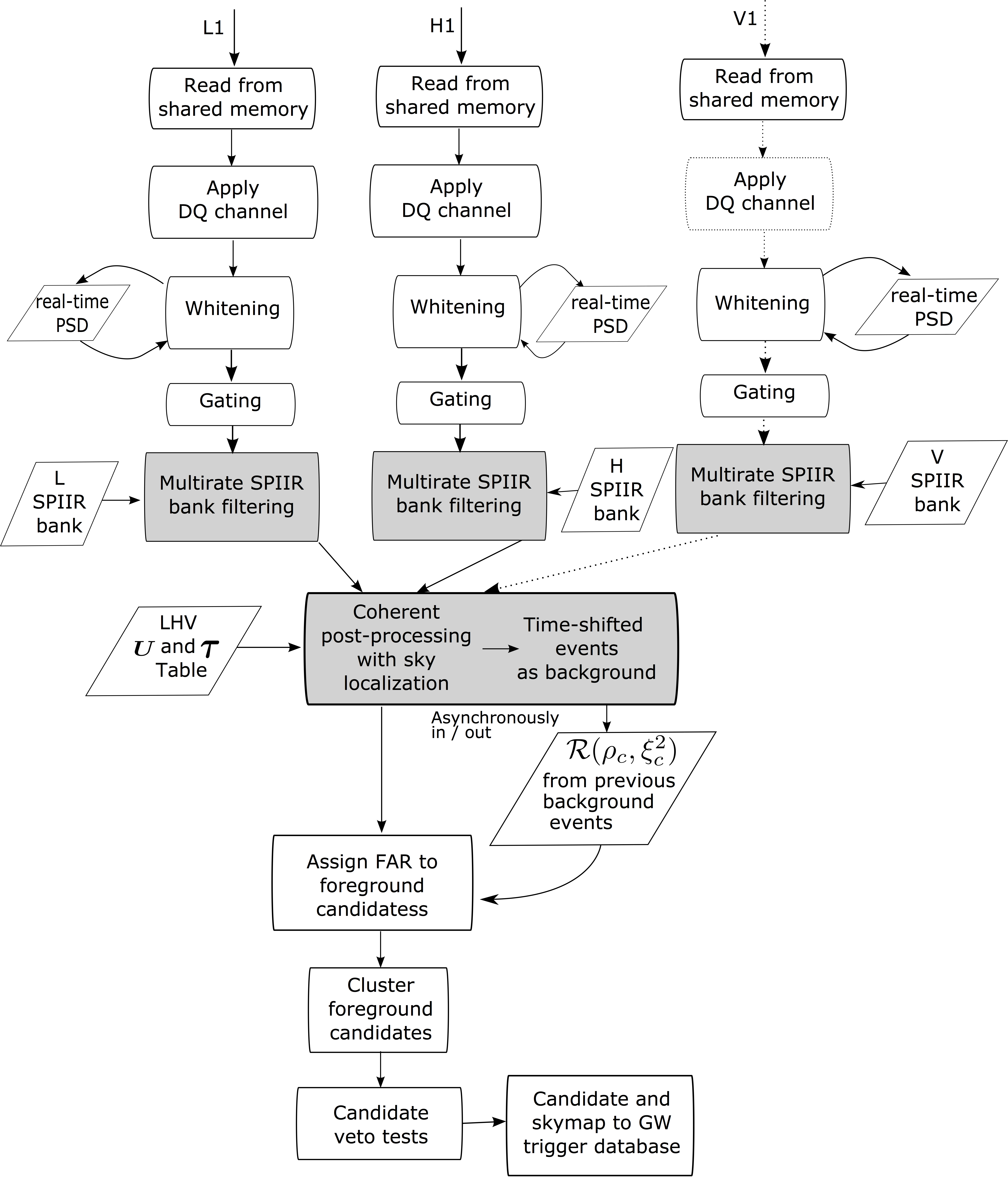}
\caption{A flowchart of the online SPIIR pipeline. Rectangular blocks denote pipeline components; those in gray have been accelerated by GPUs. Parallelogram blocks denote input for given components. At the start, the pipeline reads live data from LIGO-Livingston (L1), LIGO-Hanford (H1), and optionally Virgo (V1) (marked by the dashed line) on the computing cluster. The pipeline applies per-detector DQ flags from the DQ channel to gate noise-affected segments in the main strain data. The data are downsampled from $16\,\mathrm{kHz}$ to $2\,\mathrm{kHz}$ and whitened based on a real-time estimate of the noise power spectral density (PSD). After whitening, a gating process vetoes noise outliers. The gated data are then filtered using SPIIR filters (Sec.~\ref{sec:spiir_filtering}) and a coherent search is performed (Sec.~\ref{sec:coh_search}). An asynchronous process collects background events from time shifted data to update the distribution of the ranking statistic $\mc{R}$ from which the FAR of a foreground candidate is estimated. Candidates are then clustered and tested against thresholds before submitted along with the sky map of coherent statistics to the GW candidate database (GraceDB).
}
\label{fig:flowchart}

\end{figure}

A flowchart of the SPIIR pipeline is shown in Fig.~\ref{fig:flowchart}. The elements of the flowchart may be grouped into five stages:
\begin{enumerate}[label=(\Alph*)]
    \item A pre-processing stage, where live data streams from detectors LIGO-Livingston (L1), LIGO-Hanford (H1), and Virgo (V1) are read, conditioned using data quality (DQ) channels, downsampled, whitened, and conditioned again.
    \item A filtering stage, where whitened data are convolved with SPIIR filters in the time domain to approximate the matched filtering result -- SNR time series.
    \item A coherent search stage, where candidates from individual detectors are searched over companion detectors to form coherent candidates.
    \item Candidate significance estimation, where background (noise) events generated by time-shifted data are used to estimate the false alarm rate (FAR) of a candidate.
    \item Candidate veto and submission, where candidates are tested against a few statistic thresholds and submitted to the GW candidate event database (GraceDB).
\end{enumerate}

\subsection{Data conditioning, down-sampling, and whitening}

\label{sec:condition_whiten}
The first stage of the SPIIR pipeline uses the data acquisition module from the \texttt{gstlal} software package\footnote{\url{https://git.ligo.org/lscsoft/gstlal}} to read in live (optionally offline) data. During O3, the live data were delivered in one-second packets. The first step of the pipeline is to ``Apply DQ channel" as shown in Fig.~\ref{fig:flowchart}. The DQ channel uses bit masks to mark the periods of the strain data when not in the lock condition or affected by known loud noises. The pipeline replaces the strain data in these periods with zeros (a.k.a. gating) with a tapering filter. This zeroing method could introduce power spectral leakage in the frequency domain. The data is whitened later in the frequency domain by the pipeline where this leakage will be carried over, which could result in a slightly inaccurate whitening outcome. The leakage can be avoided if data is whitened in the time domain or by adoption of an in-painting method instead of the zeroing method shown in the recent work~\cite{zackay2019psd}. Both methods will be considered for the pipeline for the next observation run.

The strain data are then downsampled from the original $16384\mathrm{Hz}$ to a lower rate of $2048\,\mathrm{Hz}$ to reduce the computational cost downstream. This reduced rate is sufficient to capture CBC signals within the most sensitive band (15-1000 Hz) of the LIGO-Virgo detectors. The pipeline uses the \texttt{resample} module from the \texttt{gstreamer}\footnote{\url{https://gstreamer.freedesktop.org/}} library for down-sampling, which implements the method in time-domain using a finite impulse response (FIR) filter.

The data are then whitened (refers to whitening in Fig.~\ref{fig:flowchart}) in the frequency domain that can be expressed as:
\begin{equation}
     d_w(t)= \int_{-\infty}^{\infty} \frac{\tilde{d}(f)}{\sqrt{S_n(|f|)}}e^{i2\pi ft},
  \label{eq:whiten}
\end{equation} where $\tilde{d}(f)$ is data in the frequency domain obtained by Fourier transform and $d_w(t)$ denotes the whitened data. $S_n(|f|)$ is the one-sided noise power spectra density (PSD) defined through ensemble average $E()$ of the noise spectrum $ E(\tilde{n}(f)\tilde{n}^*(f')) = 1/2 S_n(f)\delta(f-f'), f>0 $. In an ideal world where the noise is stationary, the past data can be used for noise PSD estimation to whiten the current segment of data.  However the noise is known to be non-stationary in the LIGO-Virgo online data. The noise PSD is therefore estimated by tracking the geometric median PSD from overlapping 4-second blocks of immediate past data spanning 56 seconds. This estimation converges quickly and is robust against glitches (See Sec. II. B of ~\cite{gstlal_paper}). It is implemented by the \texttt{lal\_whiten} module from the \texttt{gstlal} library.

The pipeline applies a second ``gating'' function after whitening, where outstanding amplitude excursions of the whitened data are replaced with zeros. This is to remove glitches that are not identified by the online DQ channels. A empirical window of 0.25-second of zero values is padded at each side of the gated segment to suppress excess power from the glitch. This step would not lend the spectral leakage problem as following steps are performed in the time domain.

\subsection{Time-domain matched filtering with SPIIR filters}
\label{sec:spiir_filtering}
\subsubsection{SPIIR method}

The matched filtering method is the optimal method to search for known signals from Gaussian noise. It has been used in CBC searches as the CBC GW waveform templates can either be derived from post-Newtonian perturbation theories or from numerical solutions of Einstein's field equations~\cite{spintaylorf2,eob2017,imphenomD}. The parameters of the CBC waveforms can be divided into two sets, the intrinsic and the extrinsic parameter sets. The intrinsic set is related to the intrinsic properties of the GW sources --- masses and spins. The extrinsic parameters include the distance $l$, the source location ($\alpha, \delta$), the inclination angle $\iota$ and the polarization angle $\psi$, the phase $\phi_c$ and the time $t_c$ at the coalescence.  The extrinsic parameters can be represented by two parameters effectively --- the effective distance $l_{\text{eff}}$ and the termination phase $\phi_0$~\cite{bruce05}. The CBC signal $h(t)$ can then be expressed as:
\begin{equation}
h(t) =\frac{l}{l_{\text{eff}}} A(t) \cos \left ( \phi(t) + \phi_0 \right),
\end{equation}
where $A(t)$ and $\phi(t)$ are the amplitude and phase evolution respectively. The effective distance is a function of the antenna responses to the two GW polarizations ($F^{+,\times}$) and the inclination angle:
\begin{equation}
l_{\text{eff}} = \frac{l}{\sqrt{F^{+2}(1+\cos^2 \iota)^2/4 + F^{{\times}2} (\cos \iota)^2}},
\label{eq:eff_dist}
\end{equation}
and $\phi_0$ is the coalescence phase with modulation from the antenna response functions:
\begin{equation}
\phi_0 = \phi_c - \arctan \left( \frac{F^{\times}(2\cos \iota)}{F^+(1+\cos^2\iota)}\right).
\end{equation}

To search for the unknown phase $\phi_0$, a common way is to use a two-phase matched filter with orthogonal phases. This can be implemented as a complex filter ~\cite{bruce05}:
\begin{equation}
h_T(t) = h(t, \phi_0 = 0) + i h(t, \phi_0 = \pi/2).
\end{equation}
This filter is then whitened and normalized by its expected value at an effective distance of $1$ Mpc, denoted by $h_{wT}$. Using $h_{wT}$ to cross-correlate the whitened data, one obtains the matched filtering result, the complex SNR time series $z(t)$:
\begin{equation}
z (t) = \int_{0}^{\infty} d_w (\tau)  h_{wT} (t+\tau) \mathrm{d}\tau.
\label{eq:rho_w}
\end{equation}
The detected phase $\phi_0$, the coalescence time, and the effective distance can then be obtained from the detected SNR.

The SPIIR method uses a chain of first-order impulse response (IIR) filters to approximate $h_{wT}$.  Each IIR filter is consist of three coefficients to approximate a small segment of the waveform: the feedforward coefficient $b_{1}$, the feedback coefficient $a_{0}$ and the delay of the filter $t_d$. The summation of responses from all IIR filters is the approximated waveform denoted by $u(t)$. The SPIIR method is applicable to approximation of any analytical or numerical binary waveforms~\cite{jing, spiir_shaun,david_thesis}. The metric to measure the approximation accuracy is the overlap $O$:
\begin{equation}
O = \frac{(h_{wT}, u)} {\sqrt{ (u,u)} \sqrt{(h_{wT}, h_{wT})}},
\label{eq:overlap}
\end{equation}
where $(,)$ is the inner product\footnote{https://en.wikipedia.org/wiki/Dot\_product}. We optimize the coefficients such that the overlap is over 99\% which corresponds to a SNR loss of less than 1\% (see Sec.~\ref{sec:setup} however for a realistic scenario). The complex SNR from the SPIIR filtering with $N_{(m)}$ number of filters for a given template can be expressed in a discrete form as:
\begin{equation}
  z[k] =  \sum_{(m)=0}^{N_{(m)}} \left(a_{0}^{(m)} z[k-1-t_d^{(m)}/{\Delta t}] + b_{1}^{(m)} d_w[k-t_d^{(m)}/{\Delta t}]\right), 
 \label{eq:zk}
\end{equation}
where $k$ is the discrete time, $\Delta t$ is the interval of time samples.

%\begin{equation}
 %   u(t) = \Sigma_m b_{1,m} a_{0,m}^{t-t_m}.
%\end{equation}
%$b_{1,m}$ is the representative amplitude and phase of the segment, while $a_{0,m}$ has the form of $e^{\gamma + i\omega}$ to approximate the phase evolution using a linear Taylor expansion. $t_{m-1}$ is the place where the difference between the Taylor expansion and the true waveform becomes non-negligible. Here we set the allowable difference ratio for each segment to be 1\%. So the overall approximation accuracy is less than 1\%. In reality, this condition can be easily satisfied when approximating low-mass templates. High-mass templates vary sharply near the merger phase, more filters need to be placed to satisfy this condition.

\subsubsection{Computational cost and GPU acceleration}

A total of 12 floating point operations are required to calculate the SNR with one IIR filter in Eq.~\ref{eq:zk} ~\cite{jing,spiir_shaun}. The total computational cost of the SPIIR filtering in a search is proportional to the number of filters over all search templates.  Denote the average number of filters per template by $N_{(m)}$, the number of waveform templates by $N_T$, the sample rate by $N_R$ and the number of detectors by $N_d$, the computational cost for SPIIR filtering each second is then $\mc{O}(12 N_{(m)} * N_T *N_{R} *N_d)$.

A total of $412\,000$ templates were used by SPIIR during O3 covering the source component mass of $1.1 - 100~M_{\odot}$. In a typical setting of O3 where $N_{(m)}$ is around 350, $N_R$ is 2048 Hz and there are 3 detectors, the total computation is about $10.6$ Tera floating point operations per second (TFLOPS), requiring 2200 typical 4.8-GFLOPS central processing unit (CPU) cores to process in real-time.

The high demand on CPUs can be mitigated by use of GPUs. The filtering process is a multiple instruction single data process. Filtering operations of templates and of SPIIR filters are independent that they can be distributed in parallel to GPU threads. It is shown that with a moderate GPU, a speed-up of more than 100 over a single-core CPU can be achieved~\cite{liuyuan,xiaoyang2018,xiangyu2018}. The speed-up can easily scale up with more GPU cores. The frequent release of new massive-core GPU hardwares is likely to accommodate the increased computational demand due to increased detector sensitivity.

\subsection{Coherent trigger generation and localization}
\label{sec:coh_search}
\subsubsection{Coherent network SNR}
\label{sec:def_coh_snr}

The coincidence search method has long been used to search for event candidates from a detector network where high SNR triggers from individual detectors are selected first and those coincident in time are selected as candidates. The coherent search on the other hand will look for both time and phase coherent triggers from individual detectors based on the maximum network log likelihood ratio (LLR) principle. Previous work can be found in ~\cite{pai01,bose2011,harry2011,macleod2016} and Appx.~\ref{appx:llr} of this paper shows the network LLR derivation can be simplified mathematically using SVD. The coherent statistic from the network LLR is referred to the coherent SNR throughout the paper and is expressed with the SVD form as:
\begin{equation}
\rho_{\textsc{\tiny C}}^2 (\alpha, \delta, t_c, \bs{\Theta}) 
   =  \left\lVert  \bs{I}\bs{U}^T \left (
\begin{array}{cccc}
z_1 & 0 & \ldots & 0\\
0 & z_2 & \ldots & 0\\	
\vdots & \vdots & \ddots & \vdots \\
0 & 0 & \ldots & z_{N_d} \\
\end{array}
\right ) 
\right\rVert^2,\label{eq:llr_form}
\end{equation}
where $\rho_{\textsc{\tiny C}}^2$ is the coherent SNR; $\parallel \parallel$ is the Euclidean norm; $\bs{I}$ is a diagonal matrix with the first two elements non-zero $\bs{I} = \text{diag}\{1,1,0,...,0\}$, and $z$ is the complex SNR of each detector offset by different arrival times of a signal, and $N_d$ is the number of detectors in the network. $\bs{U}$ is an $N_d\times N_d$ unitary matrix from the SVD of the noise-weighted detector response $\bs{G}_{\sigma}$ given in Eq.~\ref{eq:svd}. The coherent SNR dependent on three sets of unknown parameters: the intrinsic source parameters $\bs{\Theta}$, the sky direction ($\alpha, \delta$), and the coalescence time $t_c$.

The first two columns of $\bs{U}$, spans a plane to capture the two signal polarizations from $N_d$ detectors. The complementary statistic, the null SNR which is from the null-space spanned by the remaining $N_d -2$ columns of $\bs{U}$, should only include noise contributions. It can be written as:
\begin{equation}
\label{eq:null_def}
\rho_{\textsc{\tiny NULL}}^2 = \sum_I  \rho_I  ^2 - \rho_{\textsc{\tiny C}}^2.
\end{equation}
where $\rho_I$ is the absolute value SNR $\rho_I = |z_I|$.  In the true direction, the per-detector signal arrival times and signal phases are coherently matched, the projected signals should preserve the entire signal power and the projection for the null SNR should only retains noise. In other directions where times and phases are not perfectly matched, the coherent SNR will not preserve all signal power, instead some power will leak to the null space. This lends to a localization method that explores the phase information and is expected to be better than the simple triangularization method using arrival time information.

Before we perform coherent searches, we first select candidates from each detector. A candidate is selected if its SNR is over 4 ( as used by the GstLAL pipeline~\cite{gstlal_paper}). The threshold is less than the fiducial detection threshold of 8 to allow for sub-threshold trigger and background formation. The candidates are grouped by every template bank (A template bank is a set of around $10^3$ templates grouped by chirp mass values for computational convenience.) and every second, and are filtered with the following procedure: 
\begin{itemize}
	\item Find the maximum SNR across templates of the template bank at each time sample and select those with SNRs over 4. 
	\item If a template triggers multiple high SNRs within one second, only select the candidate with the highest SNR. % s selected many times Cluster the trigger candidates in time, by picking up the highest SNR in a one-second window.
		%Mathematically, this is: $\{\hat{t_c}, \hat{\bs{\Theta}} \vert \underset{\hat{t_c} = \max{t_c}} {\rho_I}(t_{c}, \hat{\bs{\Theta}}), t_{c} \in [1s]\}_I$ %For one second of filtering results of a template, we only seFurther reduce the number of candidates in one second. only recording the the candidate that has the maximum SNR of a templat
\end{itemize}
This procedure will ensure that at each sample time, there is at most one candidate and a template will report at most one candidate during each second.

For each candidate, we then search for maximum coherent SNR using SNR time series from companion detectors. To search $(\alpha,\delta)$, we use HEALPix~\cite{gorski2005healpix} to divide the whole sky into 3072 equal-area curvilinear tiles (a.k.a. pixels). The centre coordinates for each tile are used as sky direction samples.  As Eq.~\ref{eq:llr_form} of coherent SNR can be considered as a linear projection process, the ratio of the coherent SNR from a sub-optimal location with that from the optimal direction can be modeled as $cos (\theta)$ where $\theta$ is the angular distance between this location and the optimal location. The worst scenario is when the signal lies at the boundary of a sky tile.  The maximum angular distance between any sky tile boundary and its center is $1.9^{o}$ thus the maximum SNR loss is $1-cos (\theta) = 0.05\%$. For more accurate sky tiling, readers are referred to~\cite{pygrb, harry2011,macleod2016}. 

$\bs{U}$ used for the coherent SNR calculation depends on not only the relative source sky direction but also detector sensitivities. We assume that the noise in one detector does not change much during an observing run so that a representative horizon distance is used for the sensitivity representation. To save run-time calculation of $\bs{U}$, we sample $\bs{U}$ every half hour up to 24 hours and refreshed every day to accommodate the change of detector locations due to Earth's rotation. Time offsets of SNRs from different detectors are sampled at the same interval along with $\bs{U}$. The pipeline will calculate the coherent SNR using the $\bs{U}$ and offset sample that are closest in the time to the triggering candidate. In this implementation of approximation of $\bs{U}$, the maximum coherent SNR value would not be affected but the detected sky direction for that coherent SNR value could be off by the true direction up to 15-min in the East-West direction. %The pipeline will generate the all-sky map of coherent SNR values for search of the best coherent SNR candidate but also to display directions where signals are likely to come from. %A single-detector trigger will choose $\bs{U}$ and the arrival time difference matrix closest to its $\hat{t_c}$ from the lookup tables. 

\subsubsection{Signal consistency statistic}
\label{sec:def_xi}
In addition, the pipeline uses another statistic~\cite{gstlal_paper,brucechisq} to test the consistency of the detected SNR series from each detector with expectation in time domain. The expected SNR series is projected from the autocorrelation of a template. The discrete form of the statistic $\xi_I^2$ for detector $I$ is given by:
\begin{equation}
\xi_I^2 = \frac{ \sum_{j=-N_j}^{N_j} |z_I[j] - z_I[0] A_I[j]|^2} {\sum_{j=-N_j}^{N_j} (2 - 2|A_I[j]|^2)},
\end{equation}
where $A_I[j]$ is the correlation function of a template with itself and $N_j$ is the number of time samples for comparison. The numerator is summation of a group of $\chi^2$ statistics and the denominator is the overall degree of freedom. The fraction is then a reduced $\chi^2$ with a mean value of $1$~\cite{gstlal_paper}. 

We also compute the average value of $\xi_I^2$, denoted as $\xi^2_{\text{C}}$:
\begin{equation}
\xi^2_{\text{C}} = \frac{1}{N_d}\sum_{I} \xi_I^2.
\end{equation}
This average is used along with the coherent SNR to rank coherent candidate events.

\subsubsection{Background events from time-shifting}
\label{sec:def_bg}
To estimate the significance of coherent candidates, incoherent events are constructed as background events. This is done by applying the conventional time shift technique. For every SNR>=4 candidate from any detector, we apply time shifts on other detector SNR time series with sufficiently large time offsets and use them as the background data.  The minimum offset is 0.1~seconds, which is much longer than the GW travel time between any two detectors. The number of time shifts is chosen to be 100, as limited by the GPU memory to store past data for the statistics. This sets the lower limit of our FAR from the 100 time shifts of one-week data to be around $0.5/\mathrm{yr}$. FAR values beyond this limit are extrapolated using K-nearest-neighbour (KNN) techniques shown later.

\subsubsection{Computational cost and GPU acceleration}

The computational cost for coherent search and background collection using Eq.~\ref{eq:llr_form} is estimated here. For each candidate, $\mc{O}(4 N_d^2)$ floating point operations are needed to compute the coherent SNR and the null SNR with a pre-calculated look-up table for each sky direction.  Here 4 is account for one complex multiplication and one complex addition. $N_d$ is the number of detectors. The look-up tables are prepared every day and the computing of them are negligible. The maximum number of candidates per detector per second is the number of templates $N_T$. The maximum total FLOPS is therefore $\mc{O}(4  N_d^3 N_T N_p)$ where $N_p$ is the number of sky locations, For computational efficiency, the number of sky areas searched for background events is reduced to 768 which corresponds to a SNR loss of $1\%$ compared to the optimal SNR. Note we perform 100 times calculation for the background than the foreground. The computation of $\xi^2$ for each trigger is negligible compared to the coherent network SNR computing. For a typical case of $412\,000$ templates as in O3 and three detectors, the worst-case cost is about 3.5-TFLOPS including both foreground and background calculations, comparable to the cost of the SPIIR filtering. However, in practice the number of candidates is usually one or two orders of magnitudes smaller than $N_T$. Therefore the computational cost of coherent search could be significantly less than the SPIIR filtering. The use of GPUs to accelerate this stage is described in~\cite{xiaoyang2018}.

\subsection{Ranking statistic and false alarm rate estimation}

\subsubsection{Ranking statistic and false alarm rate estimation}
\label{sec:ranking}
A coherent trigger is selected based on a list of statistics: $\{\rho_1, \xi_1^2, \rho_2, \xi_1^2, ... \rho_{N_d}, \xi_{N_d}^2, \rho_C, \rho_{\textsc{\tiny NULL}}, \xi_C^2\}$. The current coherent trigger selection criteria is to require that the SNR contribution from non-triggering detectors is reasonably significant so that $\rho_C \ge \rho_{\textsc{\tiny I}} + \sqrt{2}$ ($\rho_{\textsc{\tiny I}}$ is the SNR of single-detector candidates where the selection strategy is described earlier in Sec.~\ref{sec:def_coh_snr}). This requirement is expected to be removed in the future to allow single-detector candidates.

To construct the ranking statistic, we consider the two most discriminating statistics from the collected statistic list. The coherent network SNR $\rho_C$ reflects the overall signal strength in the detector network, while $\xi_C^2$ gives the average score from the signal morphology test. The coherent statistic and the null statistic are most useful in signal and noise classification when there are at least three detectors with comparable sensitivities. For a network of three detectors with one less-sensitive detector as is the case for O2 and O3, the values of coherent statistics would be close to the SNR additions from the two dominating detectors, L1 and H1. Besides adding V1 SNR contribution is adding V1 noise most of the time given the low sensitivity of V1. To deal with this, the pipeline provides the option to use the dominating detectors for the overall signal strength. $\rho_C$ in the remainder of this section would be replaced with $\sqrt{\rho_L^2+\rho_H^2}$. 

%The coherent triggers are ranked according to two criteria. First, given two events, if one has either: a larger $\rho_C$ with the same $\xi_C^2$, or the same $\rho_C$ with a smaller $\xi_C$; then it is more likely to be a GW signal and thus ranked higher. If that first criterion does not separate the events, their ranks are decided by the scarcity of their appearance which is the probability of observing higher-ranked candidates based on the first criterion. This ranking statistic provides a distance measurement of a candidate to the cluster of background events. 

The coherent candidates are ranked mathematically by integrating the background probability $P(\rho_C, \xi_C^2|n)$ as:
\begin{equation}
\mc{R}(\rho_C^{'}, \xi_C^{2'}) =  {\int_{\rho_C \ge \rho_C^{'}} \int_{\xi_C^2 \le \xi_C^{2'}} P(\rho_C, \xi_C^2|n)} \diff \xi_C^2 \diff \rho_C
\end{equation}
where $P(|)$ is the conditional probability and $n$ denotes the background events by time shifts. 

%The pipeline estimates the probability $P(\rho_C, \xi_C^{2}|n)$ using histogram from background events. The probability depends on the intrinsic parameters $\bs{\Theta}$. Different templates ($\bs{\Theta}$) can respond to detector noise differently in terms of number of triggers and the trigger statistics here. A foreground candidate can be insignificant in its collected background distribution but can be significant in other background distributions. Candidates need to be compared with background collected from same type of noise responses. There are no quantitative divisions on templates which noise responses are considered different. During O3, we group candidate and background by computing nodes each has $4\,000$ templates. 

The significance of a candidate is quantified by the candidate's false alarm probability, which is determined by calculating the cumulative distribution of the ranking statistic $\mc{R}$:
\begin{equation}
P(\mc{R}<\mc{R}^{'}|n) = \sum_{\mc{R}<\mc{R'}} P(\mc{R}|n),
\end{equation}
where $P(\mc{R}|n)$ is the discrete probability of $\mc{R}$, calculated by integrating the probability $P(\rho_C, \xi_C^2|n)$ for a given histogram bin. %The histogram setting of $\mc{R}$ is 300 number of bins with equal bin interval from -20 to 10 in the logarithmic scale.

\begin{figure}[htp]
\includegraphics[width=\columnwidth]{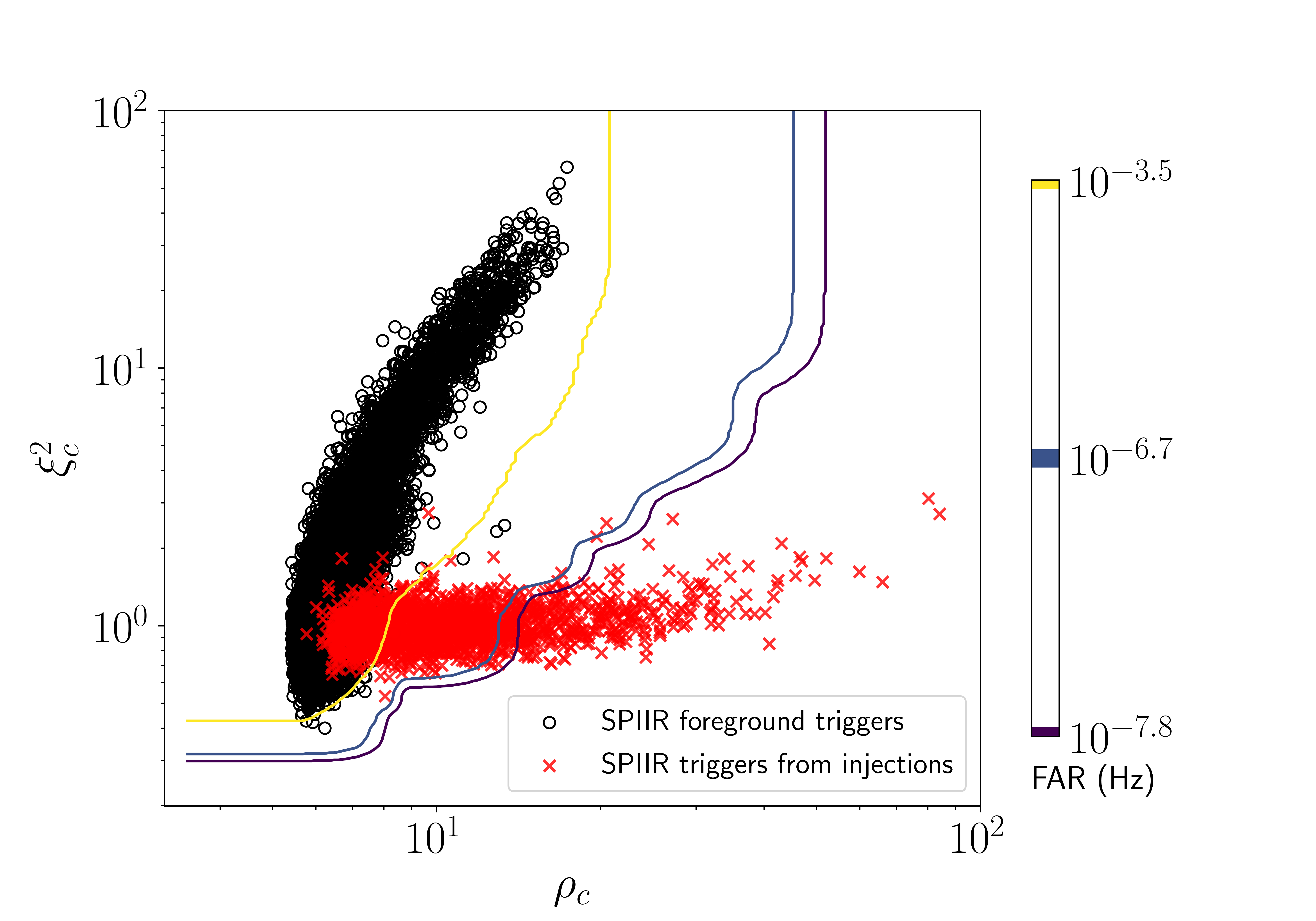}
\caption{FAR estimations using background events of one-week O2 data to separate noise triggers (black circles) and event triggers corresponding to a set of binary black hole injections (red crosses). Three representative FAR lines are shown corresponding to significances of $25.6/\mathrm{day}$ (GW database internal submission threshold), one per two months corresponding to the O3 open public alert threshold, and $0.5/\mathrm{yr}$ which is our FAR limit from data. }
\label{fig:background}
\end{figure}

The false alarm rate (FAR) is calculated by the false alarm probability and the observed trigger rate:
\begin{equation}
\text{FAR}(\mc{R}') =  \frac{P(\mc{R}<\mc{R}^{'}|n)N_b} {T_b}.
\end{equation}
where $N_b$ is the number of background events and $T_b$ is the background collection time which is equivalent to the time to collect foreground candidates multiplied by the number of time shifts performed. Apparently, different collection of background events will result in a different FAR value. During O3, we grouped background events in each computing node which processed $4\,000$ templates and adjusted the final FAR of a candidate across the total number of computing nodes. 

The pipeline collects one week of background events for FAR estimation. This sets the lower bound of FAR from data which is $~0.5/\mathrm{yr}$ satisfying the threshold of online alerts. To enable comparison of candidate significances beyond this threshold, a KNN kernel density estimation method~\cite{knn1992} was used to extrapolate the significances of SPIIR detections. An empirical $K$ value of 11 is chosen that the probability of a data bin is the Gaussian smoothed average of probabilities of the nearest 11 bins. The pipeline also collects two hours and one day of background events to capture potentially non-stationary noise behaviours in short-term and median-term . In addition, the pipeline collects individual statistics, the single SNR and the single $\xi_I^2$, from the background events and apply the same FAR estimation method for single-detector FAR values to be used in the veto stage below. The pipeline requires at least one million background events before any FAR value assignment, corresponding to a collection time of a few hours, to ensure reliable FAR estimations. Fig.~\ref{fig:background} demonstrates how the signal triggers can be separated from the noise triggers using the calculated FARs from one-week of O2 data.

\subsection{Candidate veto and submission}

A single GW signal could trigger several templates, a scenario avoided by using a clustering function at the last stage of the pipeline. Candidates are clustered based on their coherent network SNR values, within a time window, set to 0.5 seconds for O3. 
 
Once clustered, a candidate is tested to further suppress possible transient glitches. The following tests were designed to veto single-detector glitches where O1 and O2 data were used to tune the veto thresholds.
\begin{itemize}
 %   \item The candidate will be vetoed if the trigger time is within poor quality data periods marked by the data quality channels or flags.
    \item As background events were collected by three time scales of two-hour, one-day and one-week, the most conservative (maximum) FAR of the three time scales is assigned to the candidate to remove glitches that only represent in any of the short-term, median-term, or long-term scales. 
   \item  FARs estimated from at least single detectors need to be relatively significant. The single FAR threshold was tuned with O1 and O2 data and 0.5 Hz was found to be sufficient to reject all load single-detector glitches while retain all detections.
    \item We reject a trigger if there has been a submitted trigger within the last 50 seconds with a FAR value less than a factor of two of the FAR of the test trigger. This helped remove spike submissions of a real signal or periodic transient glitches.
    \item Due to that two SPIIR triggers associated with clear loud glitches were uploaded on Jan. 2020~\footnote{https://gracedb.ligo.org/superevents/S200106au/view/, https://gracedb.ligo.org/superevents/S200106av/view/} with extremely high $\xi^2$ values and alert significances, an additional test was implemented to veto candidates if any of the single $\xi_I^2$ values is more than 3.  This value is chosen to retain all O1 and O2 detections including the glitch-affected GW170817 event and is effective in glitch veto for the remaining of the O3 run.%The possibility of a signal with $\xi^2$ larger than 3 is less than 10\% in Gaussian noise. Typically for real signals, $\xi^2$ values are always close to the mean value one:

\end{itemize}

If a candidate passes all tests, it is uploaded to GraceDB with a table of the network SNR, the coherent SNR, individual SNR, $\xi_I^2$ statistics and mass and spin parameters as well as the time of the merger.

\section{Pipeline performance and O3 online run}
\label{sec:results}

\subsection{Data, pipeline, and injection setup}
\label{sec:setup}
Here we show the performance of the pipeline using the O2 open data~\cite{open_data_o1o2}. Category 1 and 2 (CAT1, CAT2) flags were used to flag periods of poor quality and will be set to zeros by our conditioning procedures. The offline open data went through several cleaning and calibration processes than the online data. The CAT1 and CAT2 data quality flags benefited from post-run measurements of noise-witness channels and provided more data quality information that are not necessarily available online~\cite{open_data_o1o2,davis2019}. The data used here span from  $1\,186\,248\,818$ GPS seconds (Aug 13, 2017 at 02:00:00 UTC) to $1\,187\,312\,718$ GPS seconds ( Aug 21, 2017, 05:00:00 UTC). The total data duration is 687900 seconds, i.e. 7.96 days. For this period of time, the two LIGO detectors were mostly in duty with a joint duty cycle of about 70\%. For about 25\% of the time one of the LIGO detectors was also joined by the Virgo detector.

The CBC templates used during the SPIIR O3 run were obtained from~\cite{gstlalbank2018}. The original templates cover binary neutron star (BNS), neutron star and black hole (NSBH), and intermediate-mass binary black hole (BBH) systems where the total masses are between 2 $\mathrm{M}_{\odot}$ and 400 $\mathrm{M}_{\odot}$ and the mass ratios are between 1.0 and 98.0. The spin parameter of the system is set to be the aligned spin on the z-component. For component mass less than 2 $\mathrm{M}_{\odot}$, the spin is set within $\pm0.05$. Otherwise, it is set to be within $\pm0.999$. For the SPIIR O3 run, the templates were down selected by restricting the component mass to be over 1.1 solar mass as a NS mass to be less than 1.1$\mathrm{M}_{\odot}$ is unlikely from estimation~\cite{neutronstar2016}. We further constrained the upper bound of the component mass to be less than 100 $\mathrm{M}_{\odot}$ by computing resource consideration. This gave $412,000$ templates which fit into 103 computing machines on the LIGO-Caltech cluster. Each computing machine is equipped with a Quad-Core AMD Opteron(tm) 2376 CPU and a Nvidia GTX 1050Ti GPU.

Two injection sets were used to test the detection performance of the pipeline, the BNS injection set and the BBH injection set. We expect the pipeline detection performance of NSBH injections to be between the performances of the two injection sets here. Injections were placed every 1000 seconds in the O2 data. The BNS injection parameters were sampled as follows: a uniform distribution for component masses with the range of 1.1 $\mathrm{M}_{\odot}$ to 2.3 $\mathrm{M}_{\odot}$, an isotropic distribution up to 0.4 for spin, a uniform distribution for sky position, and a uniform volume distribution and up to redshift 0.2 for distance. The BBH injection parameters are drawn from the same distributions for sky positions and distances respectively except the distance range was increased to a redshift of 0.7. The primary mass for the BBH injection was drawn from Salpeter IMF distribution between 5 $\mathrm{M}_{\odot}$ and 50 $\mathrm{M}_{\odot}$ and the secondary mass was drawn uniformly between 5 $\mathrm{M}_{\odot}$ and the primary mass. 

\begin{figure}
\includegraphics[width=\columnwidth]{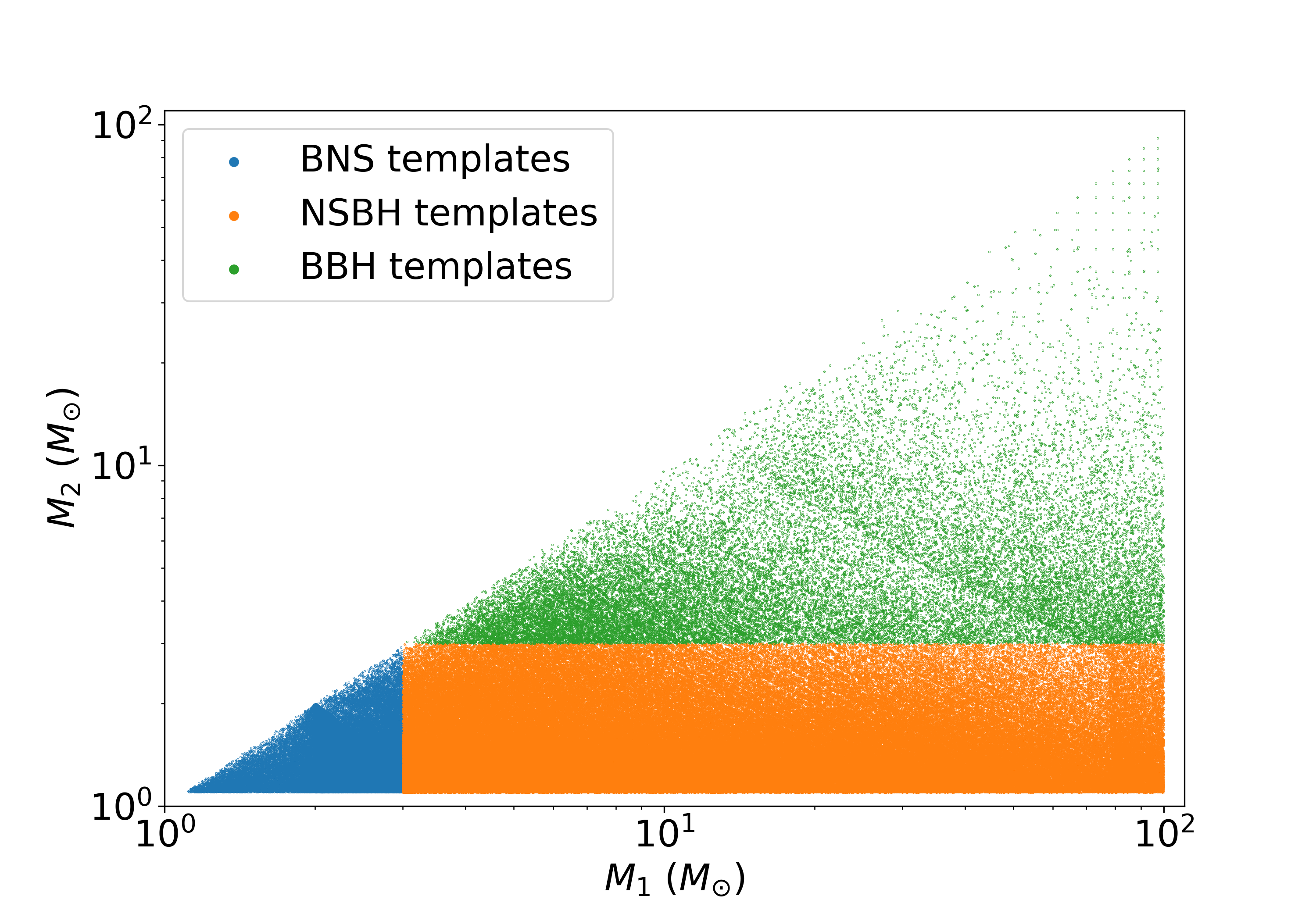}
\caption{Component mass of templates used for the O3 SPIIR online search. The templates are divided into BNS, NSBH, and BBH categories for performance demonstration using BNS and BBH injections. The BNS templates are chosen so that both component masses are less than 3 $\mathrm{M}_{\odot}$. The boundary for the BBH category is that both component masses are more than 3 $\mathrm{M}_{\odot}$. }
\label{fig:tmplts}
\end{figure}

\begin{figure}
\includegraphics[width=\columnwidth]{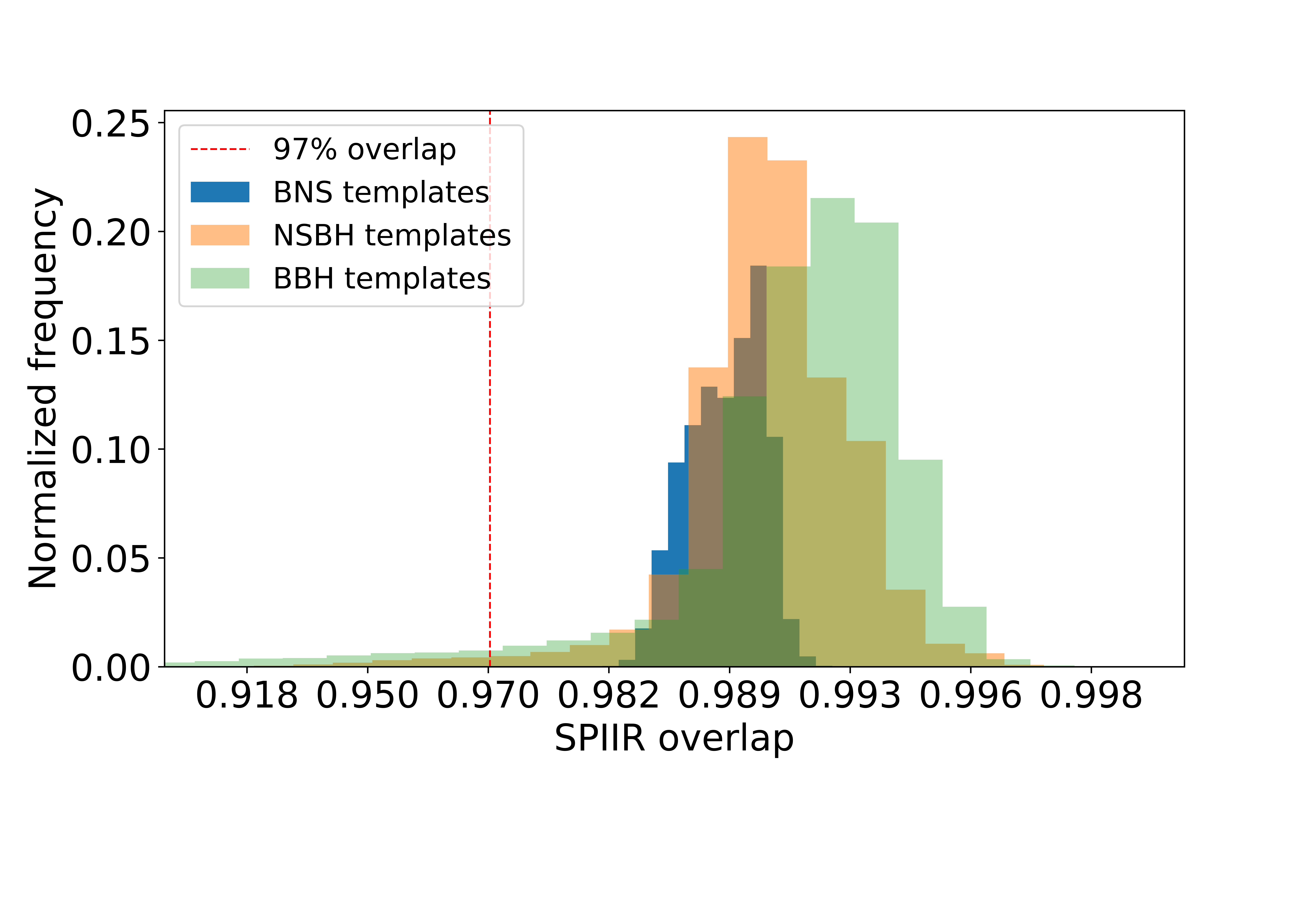}
\caption{Overlaps between the responses of SPIIR filters and the original templates for each BNS, NSBH, and BBH category. Overlap of 97\% is marked by the red dashed line.}
\label{fig:overlap}
\end{figure}

In order to demonstrate the pipeline's performance with injections, the templates were divided into three categories to detect BNS, NSBH and BBH injections  respectively (see Fig.~\ref{fig:tmplts}).  Fig.~\ref{fig:overlap} shows the overlaps (Eq.~\ref{eq:overlap}) for SPIIR approximation accuracy in each category. The majority of overlaps are more than 97\%, meaning that SNR loss would be less than 3\% from the SPIIR filtering. In general the overlaps for heavier systems are higher than that of BNS systems. This is due to the fact that we limit the number of SPIIR filters to be no more than 350. The same number of filters are used to patch shorter signals from BBH systems, yielding higher overlaps. However there are less than 5\% of the NSBH and BBH systems with overlaps less than 97\%.  For comparison, only less than 0.1\% of the BNS template overlaps are less than 97\%. This is due to large asymmetry of some binary systems causing sharp variations in signals that the limited number of filters are insufficient to capture the profile.

\begin{figure}
\includegraphics[width=\columnwidth]{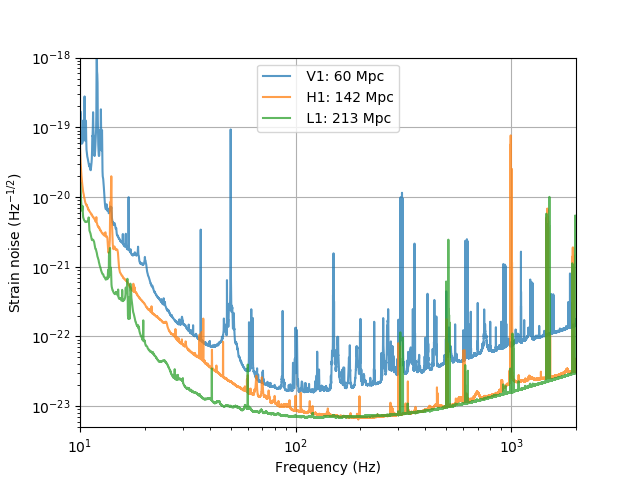}
\caption{Representative amplitude spectral density of the noise in H1, L1, and V1 detectors from Aug.13 2017 to Aug. 21 2017. The horizon distance is the maximum distance a detector can observe an optimally oriented binary source of 1.4 $\mathrm{M}_{\odot}$ and 1.4 $\mathrm{M}_{\odot}$ with the signal-to-noise ratio no less than 8.}
\label{fig:psd}
\end{figure}

The pipeline was setup to use a representative noise PSD, the median PSD over this survey period, in data whitening, template whitening, and calculation of expected SNRs from injections for simplicity. In reality, the detector noise is known to be non-stationary in very short time scales and the effect of using a different PSD than the true PSD for SNR is measured to be up to 5\% for significant events~\cite{zackay2019psd} using open data. The PSD fluctuation is expected to be larger in an online run where the pipeline needs to track PSD as mentioned in Sec.~\ref{sec:condition_whiten} for whitening. Fig.~\ref{fig:psd} shows the amplitude of the noise spectral density from the representative PSD. PSD determines the detector sensitivity for different binary systems. The conventional method to represent a single-detector's sensitivity is the BNS horizon distance computed from the PSD as shown in Fig.~\ref{fig:psd}. The pipeline used the combined SNR from H1 and L1 SNRs for ranking as explained in Sec.~\ref{sec:ranking}. The Virgo sensitivity is low compared to the LIGO detectors and therefore not used for ranking but for sky localization. To prepare background events for immediate significance estimation, the pipeline was run for a whole week before the survey start and ran uninterruptedly for the survey period.

\subsection{Injection performance}

\begin{figure*}[!htb]
\centering
\subfigure[]{
\includegraphics[width=\columnwidth]{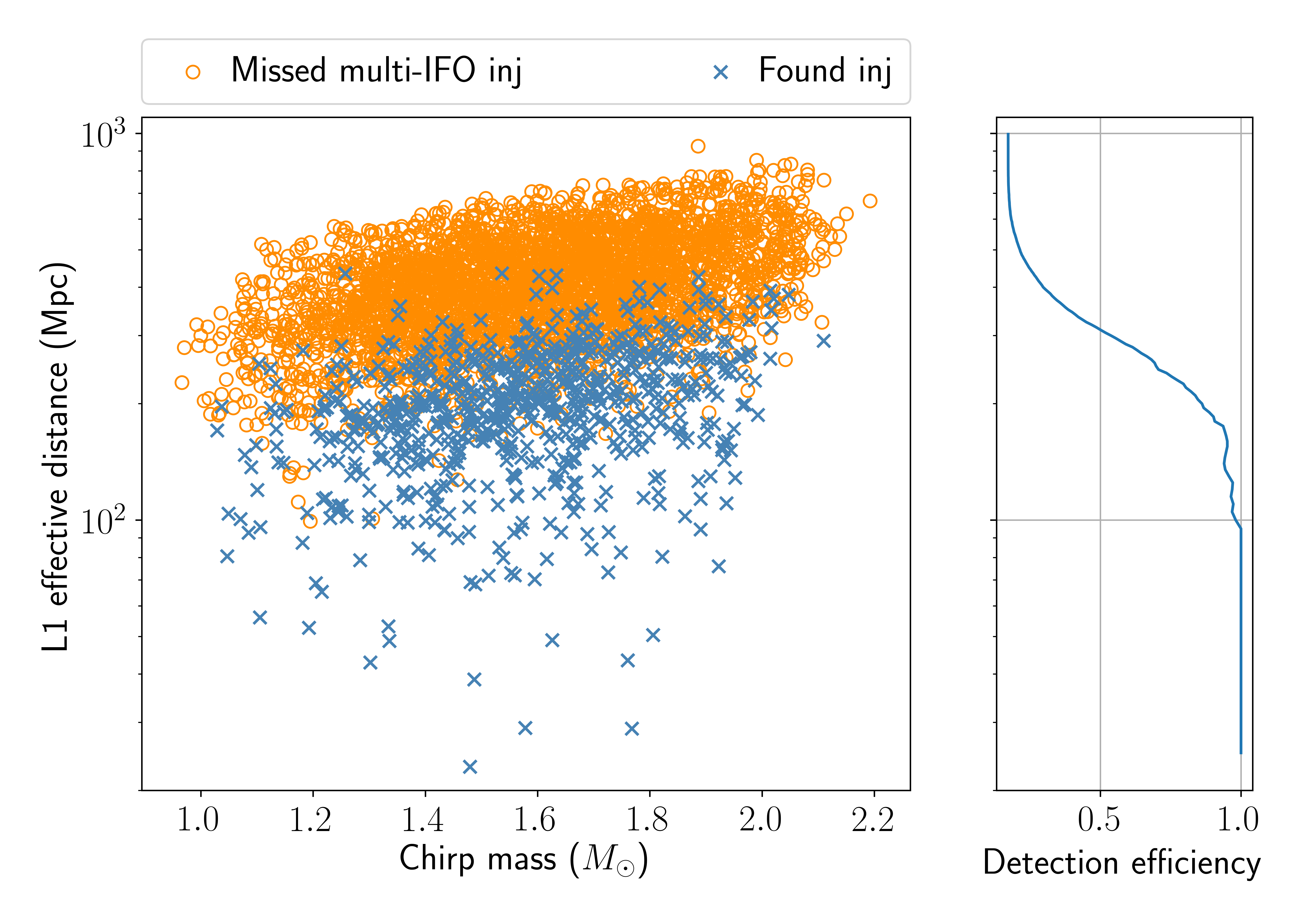}
} 
\subfigure[]{
\includegraphics[width=\columnwidth]{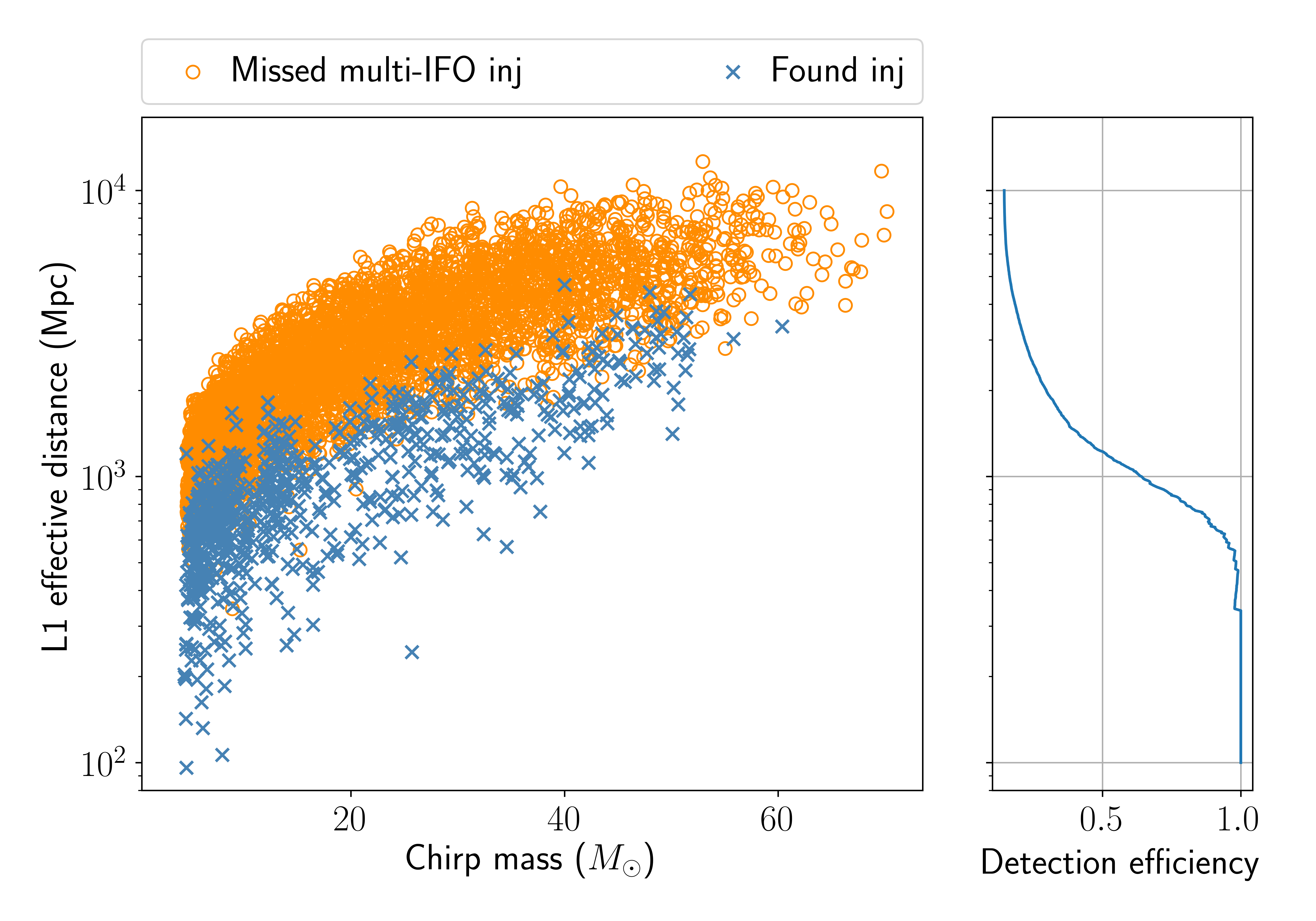}
}
\caption{(a) Missed (orange) and found (blue) BNS injections and the SPIIR detection efficiency in terms of the LIGO-Livingston (L1) effective distance and the binary chirp mass. (b) Missed and found BBH injections by the pipeline. Heavier systems that could generate stronger GWs can be detected farther away. The detection efficiency of NSBH sources are expected to between these two types of sources.}
\label{fig:inj_missfound}
\end{figure*} %\end{strip}

The performance of the injection runs is demonstrated in scatter plots (Fig.~\ref{fig:inj_missfound}) in terms of missed and found against effective distance (Eq.~\ref{eq:eff_dist}) and chirp mass. The chirp mass is a function of the individual masses and is the leading term in the signal evolution, expressed as:
\begin{equation}
\mc{M} = \frac{(m_1m_2)^{3/5}}{(m1+m2)^{1/5}},
\end{equation}
where $m_1$ and $m_2$ are component masses. An injection is considered detected if the pipeline trigger is within $\pm$0.9 seconds of the injected time and the trigger significance is better than one per two months. The most sensitive detector, L1, has been used to demonstrate the performance in Fig.~\ref{fig:inj_missfound}.

The pipeline can detect 100\% BNS events at distances less than 100 Mpc and more than 50\% of BNS events if the source distances are less than 300 Mpc. For BBH injections, the pipeline can detect 99\% sources when the source distances are less than 500 Mpc and more than 50\% of sources when the distances are less than 1000 Mpc. This is consistent with the expectation that heavier systems which generate stronger GWs could be detected farther away. Two injections within 500 Mpc were missed by the pipeline because the locations and orientations of the sources were disfavored by the H1 detector resulting a SNR below the selection threshold of 4 and our O3 pipeline was not prepared to detect GWs from one detector only.

\begin{figure*}[!htb]
\centering
\subfigure[]{
\includegraphics[width=\columnwidth]{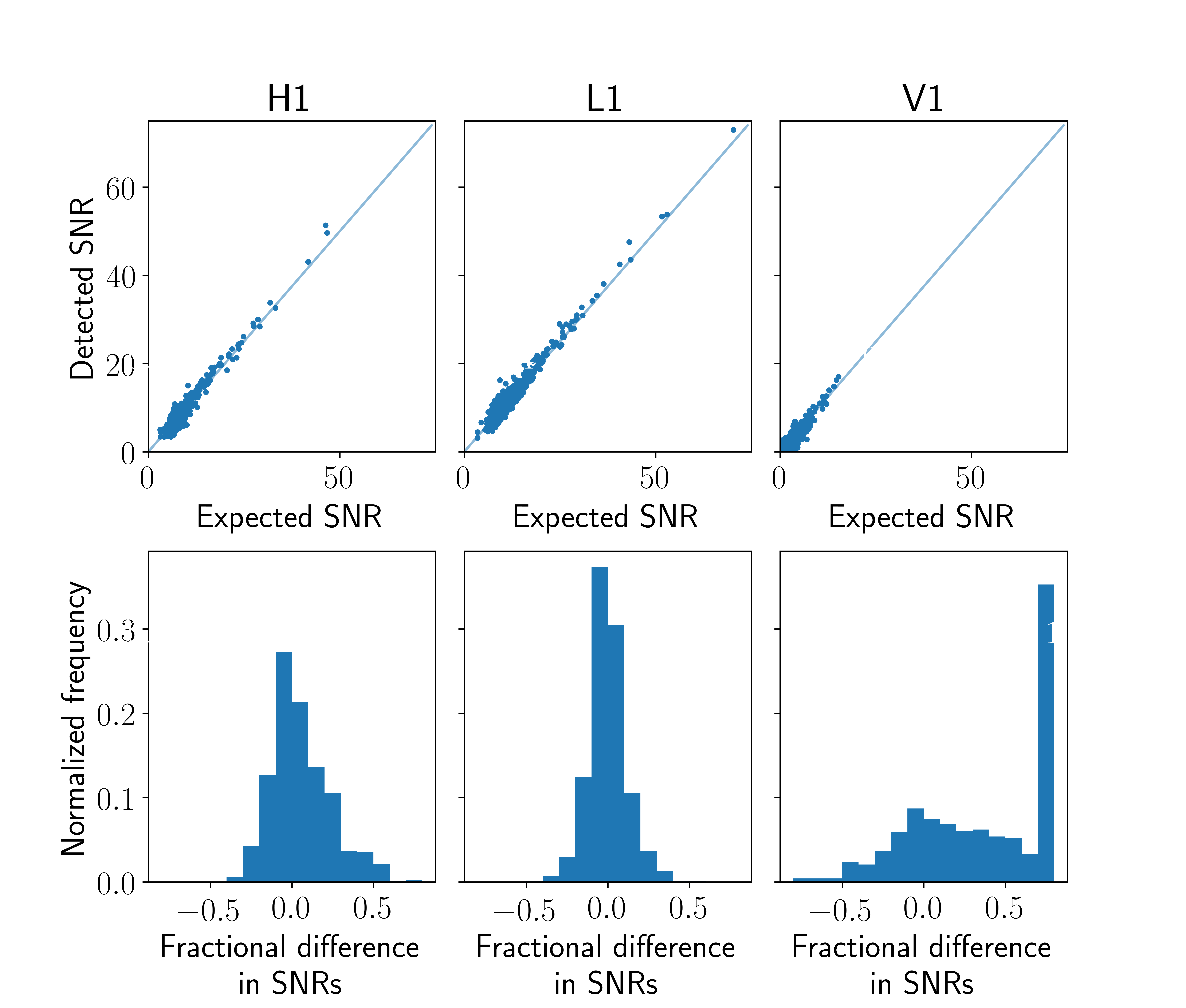}
} 
\subfigure[]{
\includegraphics[width=\columnwidth]{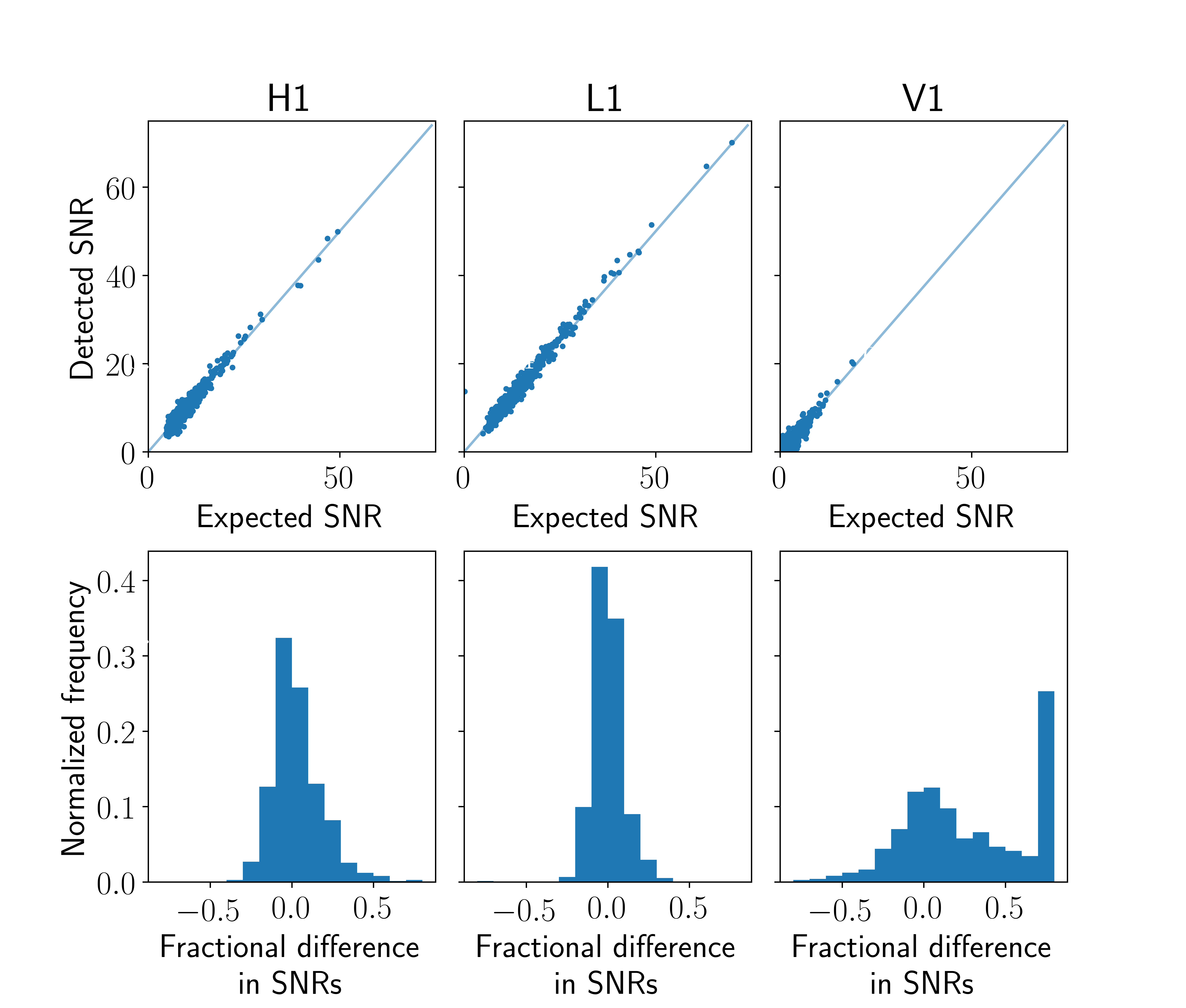}
}
\caption{Top panel: expected SNRs vs. detected individual SNRs of (a) BNS detections, (b) BBH detections using O2 data. Bottom panel: fractional difference of detected SNRs and expected SNRs from (a) BNS detections, (b) BBH detections.  SNR fractional differences are averagely around $11\%$ for H1 detections and 7\% for L1 detections. Around 40\% of V1 detected SNRs are at least 50\% off the expected SNRs as shown in the last bins of the histograms. Due to low sensitivity of Virgo in O2, most of the expected SNRs are less than 2 that the large uncertainty in SNR detection is expected. }
\label{fig:inj_snr}
\end{figure*}

Fig.~\ref{fig:inj_snr} shows the SNR recovery accuracy with H1, L1 and V1 detectors. The expected SNRs are computed from perfectly matching templates. There is an average of $11\%$ error in H1 SNR recovery and $7\%$ error in L1 SNR recovery.  Up to $5\%$ SNR deviation is expected with a fixed PSD ~\cite{zackay2019psd}. The other factors are the mismatch between the search template and the injection and the SPIIR approximation to the search templates. The large uncertainty in V1 SNR estimations is due to the low sensitivity of Virgo in O2. 40\% of V1 SNRs are expected to be less than 2 and these V1 detections would be indistinguishable from V1 noise triggers as shown by the last bin of V1 SNR error in the figure. The V1 SNR estimation should be improved along with the sensitivity improvement in O3 and beyond.

\begin{figure*}[!htb]
\centering
\subfigure[]{
\includegraphics[width=0.75\columnwidth]{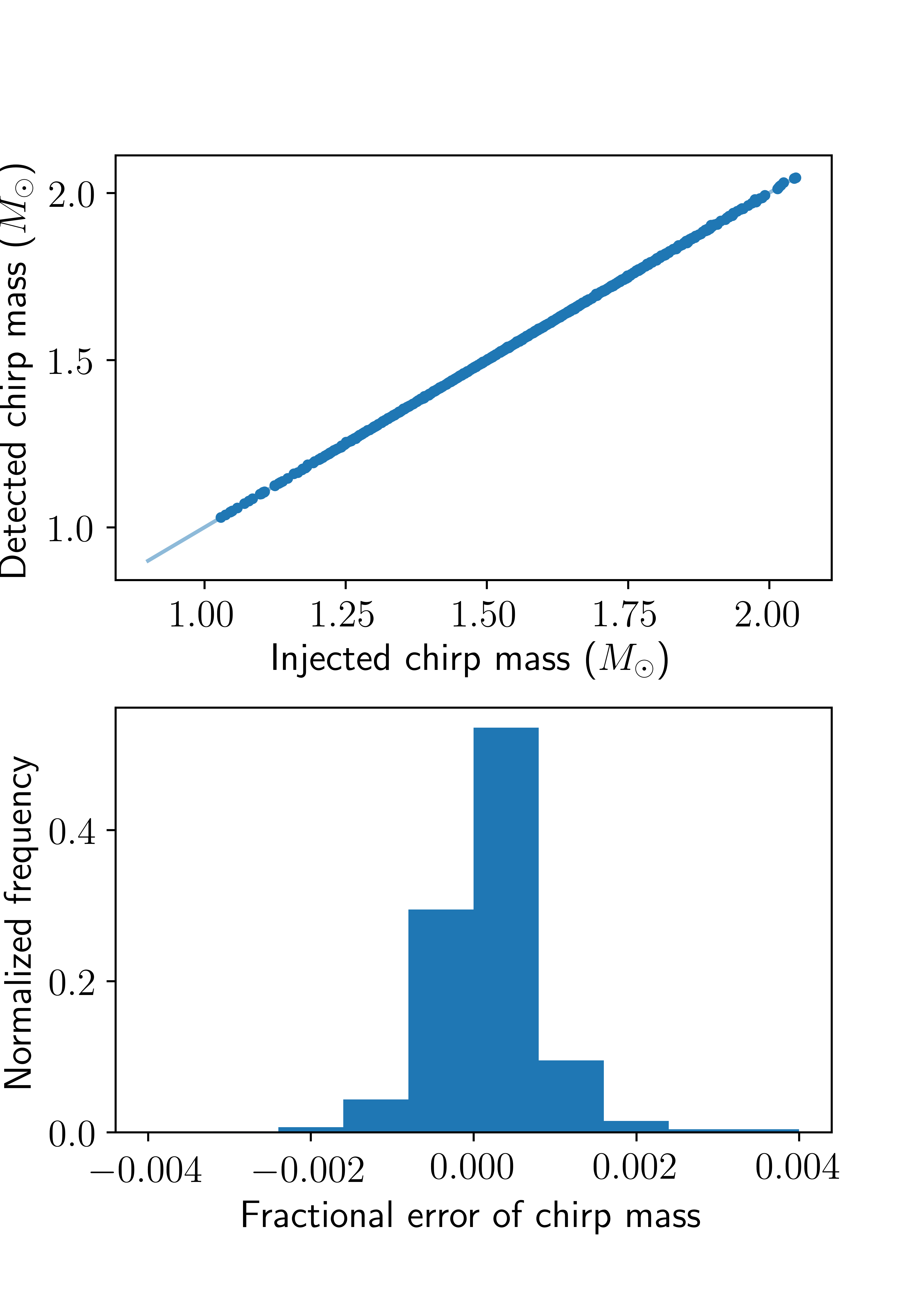}
} 
\subfigure[]{
\includegraphics[width=0.75\columnwidth]{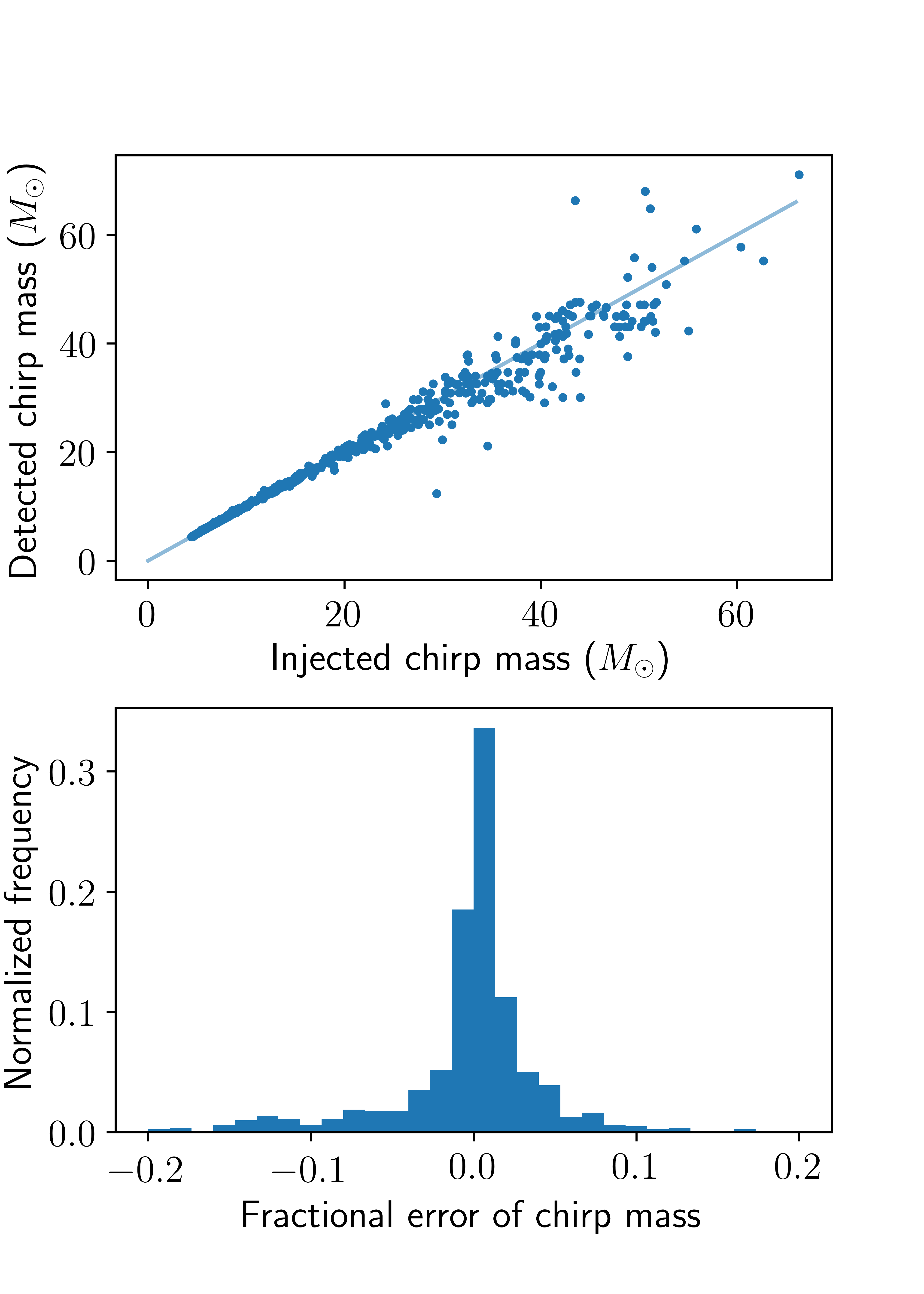}
}
\caption{Injected vs. detected chirp mass (top panel) and fractional error (bottom panel) of (a) BNS detections, (b) BBH detections. The errors for BBH chirp mass detections are larger in the high chirp mass region but are still within 20\%. }
\label{fig:inj_mchirp}
\end{figure*}

Fig.~\ref{fig:inj_mchirp} shows the detected chirp mass for BNS and BBH injections. For BNS detections, the templates were placed much denser so that the detected chirp masses are within $0.4\%$ of true values. The errors for BBH chirp mass detections are larger in the high chirp mass region. This is partly due to that the BBH templates are placed more sparsely in the high chirp mass space and partly due to that the uncertainty is larger in high mass region\cite{veitch15}.

\subsection{O2 Chunk search results}
\subsubsection{GW170814, GW170817, and GW170818}
\begin{figure}[!htb]
 \includegraphics[width=\columnwidth]{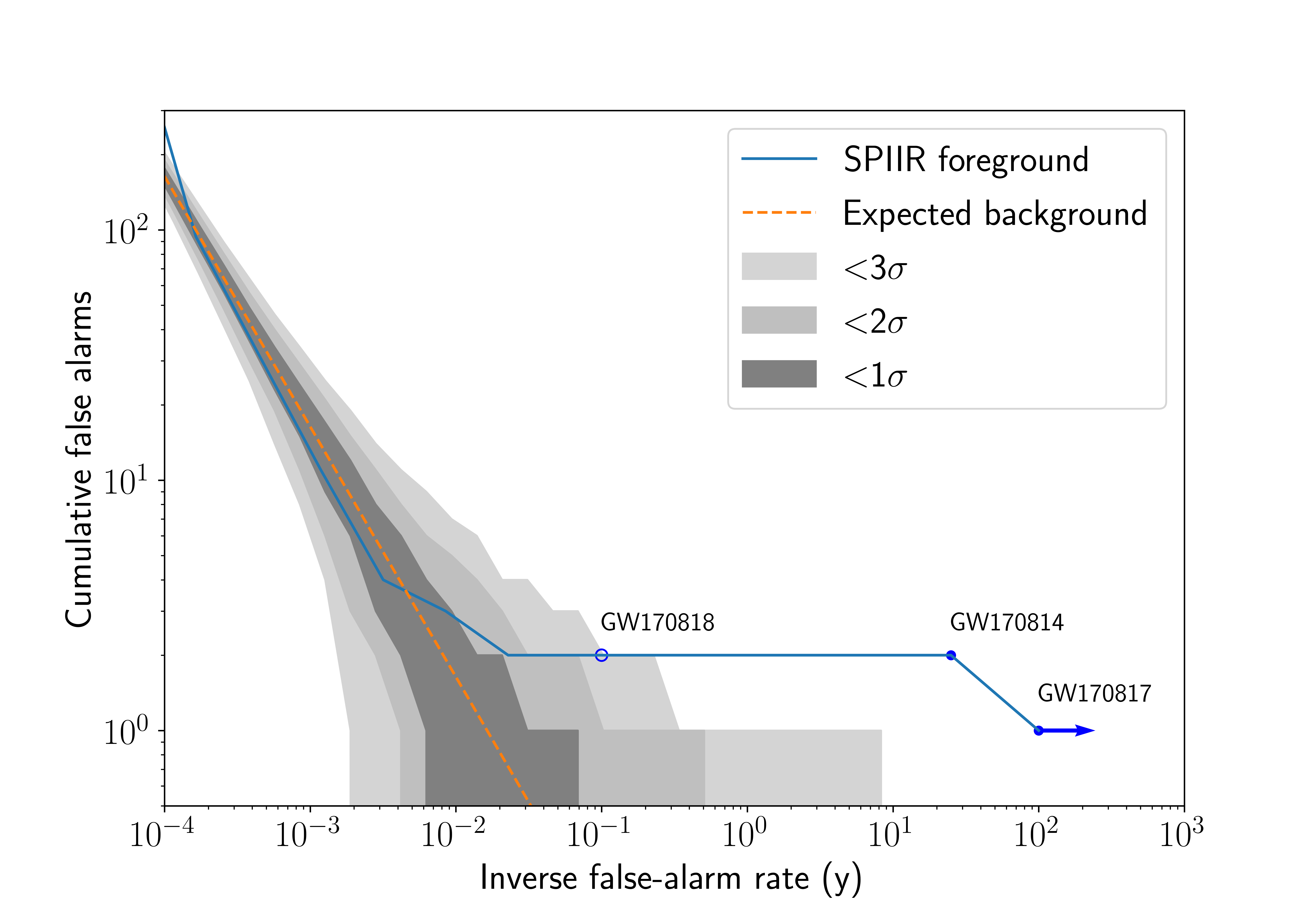}
  \caption{Search results of H1, L1, and V1 data from Aug.13 to Aug.21, 2017 using the O3 SPIIR pipeline. Orange line is the expected number of false alarms given the threshold of inverse false alarm rate (IFAR).  Shaded area shows the Gaussian standard deviations for the number of the false alarms. Blue curve is the number of SPIIR foreground candidates from this survey. The GW170818 detection has a significance of $11/\mathrm{yr}$ if removing GW170814 influence from the background.}
  \label{fig:ifar}
\end{figure}

The pipeline used the entire bank set (BNS+NSBH+BBH) to search over the same chunk of data without injections. It reported two significant triggers to events GW170817 and GW170814 as shown in Fig.~\ref{fig:ifar}. These two events were also reported online by other pipelines during the O2 run~\cite{gwtc1}.

The GW170818 event was reported in the offline GstLAL~\cite{gwtc1} and PyCBC searches~\cite{Nitz2020}. There was a GstLAL online trigger matching this event but with a marginal significance~\cite{gwtc1}. SPIIR reported a trigger with $4.4\times10^2/\mathrm{yr}$ in this search. The significance of GW170818 SPIIR trigger was affected by the precedent background from the GW170814 detection. The GW170814 and GW170818 parameters are close in parameter and time that they are in the same background collection bin. If we remove the GW170814 event trigger from the GW170818 background collection, the significance of the GW170818 trigger becomes $11/\mathrm{yr}$. In the future runs where detections are more frequent, the SPIIR online pipeline is planned to adopt strategies to remove influences of detections from background. The significance of GW170818 trigger is not affected by itself as the online pipeline only collects background from history.

Table~\ref{tab:event_tab} gives the individual, the network, the null SNRs and the significance for the three SPIIR detections. The network SNR is the quadrature sum of individual SNRs and is consistent with that of the GstLAL (GW170817 SNR 33.0, GW170814 SNR 15.9, GW170818 SNR 11.3) and PyCBC pipelines (GW170817 SNR 30.9, GW170814 SNR 16.3)~\cite{gwtc1}.The null SNRs are not used in the SPIIR pipeline for trigger ranking but are presented here for comparison to the Gaussian noise expectations. The mean value of the null SNR square from Gaussian noise should be $2N_d - 4$ which is 2 in this search. The detected null SNR values are smaller than the expectation meaning that the detections are less likely due to transient glitches. The null SNR information is not explored by the O3 pipeline as the signal null SNR and noise null would be indistinguishable with the current network configuration. This information is considered for future versions of the pipeline when there are at least three detectors with comparable high sensitivity. 

\begin{figure*}[!htb]
\centering
\subfigure[]{
\includegraphics[width=\columnwidth]{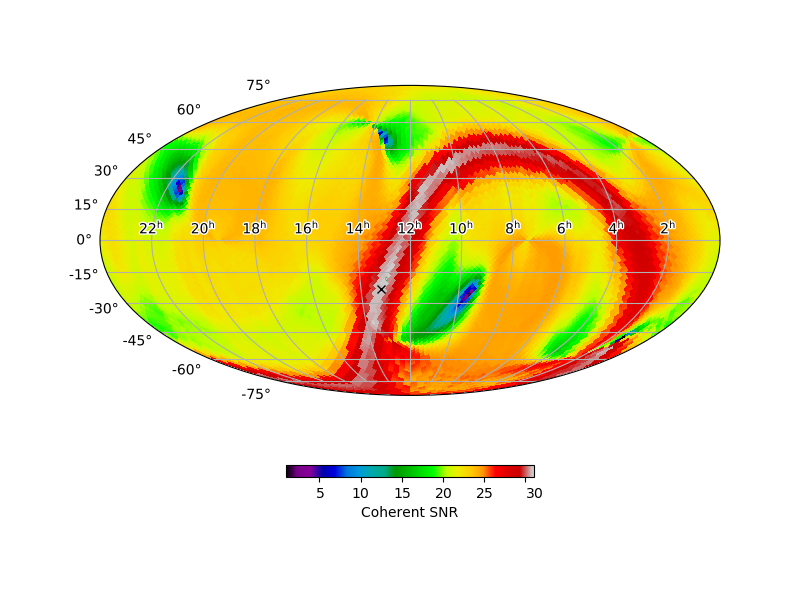}
} 
\subfigure[]{
\includegraphics[width=\columnwidth]{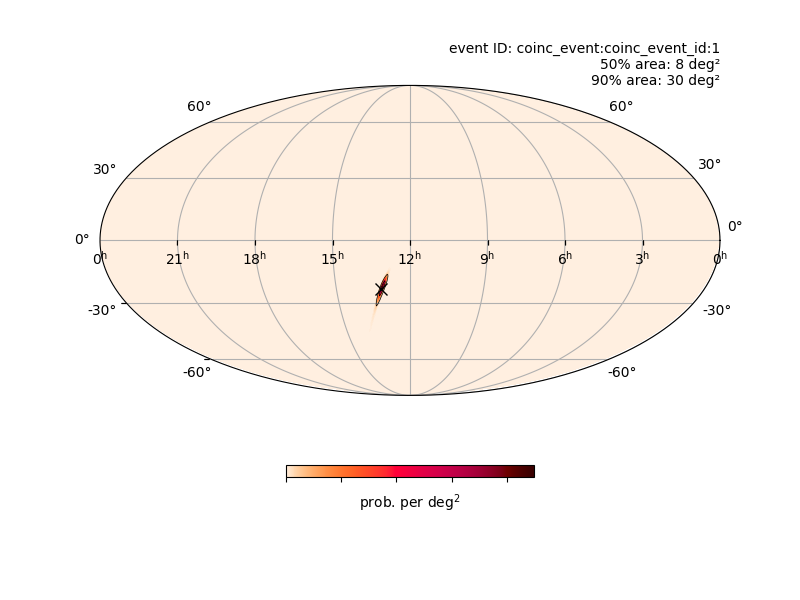}
}
\caption{\label{fig:gw170817_skymap} (a) The all-sky map of coherent SNRs searched by the SPIIR pipeline for the GW170817 detection. (b) Rapid sky localization, \texttt{Bayestar}, using the SPIIR detection. The black cross marks the sky direction from the optical discovery~\cite{gw170817_mma}.}.

%The high coherent SNR patch in the coherent SNR skymap is larger than the \texttt{Bayestar} localization area. This is due to that the coherent SNR of the four extrinsic parameters when . The \texttt{Bayestar} marginalized these extrinsic parameters breaking the degeneracy.  }
\end{figure*}

The pipeline outputs a sky map of coherent SNR values for each significant candidate. Fig.~\ref{fig:gw170817_skymap} shows such coherent SNR skymap for the most significant detection during this search, the GW170817 SPIIR detection. The optical discovery of the event is highlighted and well captured by the high SNR area. The \texttt{Bayestar} program has been used to rapidly construct source sky localizations using CBC triggers during online runs. The localization of GW170817 using the \texttt{Bayestar} program with the SPIIR detection is shown in Fig.~\ref{fig:gw170817_skymap}. The 90\% localization area from the SPIIR detection is consistent with the published localization using the \texttt{Bayestar} program with a PyCBC trigger\footnote{\url{https://dcc.ligo.org/LIGO-G1701985/public}}. Comparing the two sky maps in the figure, the coherent SNR map computes the network likelihood ratio optimized over four extrinsic parameters, while the \texttt{Bayestar} method takes prior information of all extrinsic parameters and calculates the marginalized posterior distribution. 

%degeneracy of the four extrinsic parameters that are maximized when calculating coherent SNR. The  marginalized these extrinsic parameters breaking the degeneracy.  The coherent SNR skymap is calculated maximizing the four main extrinsic parameters while \texttt{Bayestar} localization marginalizes all extrinsic parameters. The 90\% localization area from the SPIIR detection is consistent with the publicly released Bayestar localization from the PyCBC detection

%This method has been developed and will be included in the SPIIR pipeline for O4. The coherent SNR is from maximization of four nuisance parameters. By proper marginalization of these parameters, the posterior probability of location parameters can be obtained. 
%is within the high coherent SNR area. The 95\% confidence area is 30 $\text{deg}^2$, overlaps well with the \texttt{LALInference} skymap at 90\% confidence level of 28 $\text{deg}^2$~\cite{gw170817}. Due to the assumption on the flat prior and the very course sky division in the current implementation, the localization boundary is not very accurate given such strong signal. It is planned to be improved by using a realist prior and using a granular and adaptive sky division method.

\begin{table}
\begin{minipage}{\columnwidth}
    \begin{tabular}{|c|c|c|c|p{1.2cm}|c|c|}
    \hline
      Event detection & $\rho_{\textsc{\tiny L}}$ & $\rho_{\textsc{\tiny H}}$ & $\rho_{\textsc{\tiny V}}$ & Network SNR  & $\rho_{\textsc{\tiny NULL}}$ & $\mathrm{FAR}$ ($\mathrm{y}^{-1}$)\\
      \hline
      GW170814  &  13.8 & 8.9 & 3.8  &16.9 & 0.3 & $0.04$\footnote{\label{2event}$0.5/\mathrm{yr}$ is our FAR limit from data, the significant FARs here are extrapolated using KNN for the purpose of comparison.} \\
      GW170817   & 24.4 & 19.4 & 3.7 &31.5 &  1.3 & $<10^{-10}$\footnotemark[1]\\
      GW170818  & 9.9 & 4.4 & 4.1 & 11.6 & 1.0 & $4.4\times10^2$ ($11$\footnote{If GW170814 influence is excluded from background.} ) \\
    \hline
    \end{tabular}
    \caption{Individual, network, null SNRs and significance of three SPIIR detections in the search of H1, L1, and V1 data from Aug.13 to Aug.21, 2017. The expected distribution of the null SNR square in Gaussian noise is a central $\chi^2$ distribution with the degree of freedoms 2.  }
    \label{tab:event_tab}
\end{minipage}
\end{table}

% GW170817: 8.7e-27 Hz
% GW170814: 1.3e-9 Hz
% GW170818: 1.4-5Hz include GW170814 as background, 3.8-7 Hz remove the influence

%\begin{figure}
%  \includegraphics[width=\columnwidth]{i}
%\caption{IFAR plot for multiple FAR estimates from different bootstrap samples of 000 background points \todo{Chichi/Teresa}}
%\end{figure}

 \subsection{Latency}
\label{sec:latency}

The pipeline latency is defined as the time between receiving data and a trigger is produced. The latency in general comes from two sources: the intrinsic delay to collect required size of data, and the time spent on computing. Tab.~\ref{tab:latency} provides a list of main latency sources and possible future improvement.  $N_{\text{rate}}$ is the data sample rate (2048Hz for O3). At the beginning of the pipeline, the data is packed in one-second packets causing a waiting time of one second. The size of a data packet in theory can be reduced to as small as one sample. Next the downsampling module uses a FIR filter that introduces a waiting time equaling the filter length $N_{\textsc{\tiny FIR,D}}$ which has a typical length of 192. The downsampling precision is proportional to the length of this filter. We assume this filter does not change for simplicity. Following on, the whitening process collects the segments of data and performs PSD division on them in the Fourier domain. The whitening latency is dependent on the segment length which is set to two seconds in O3. A recent work proposed a time-domain whitening that can reduce the latency to be zero-second~\cite{leo_td}. The next stage gating applies a 0.25 second of side window and we simply assume the same for future runs. In the coherent search stage, it calculates $\xi_I^2$ over a series of data samples that requires a collection of $N_j$ data samples (typically 175). In the last stage of candidate clustering and veto process, a 0.5 second window is applied to cluster triggers. This could be removed by applying submission thresholds. Though data comes in one-second packets, they would be sliced and combined at different stages of the pipeline. E.g. the gating module will slice data at the boundaries of bad data thus producing irregular durations of data packets downstream while the filtering and coherent search modules only work on integer seconds of data that they might wait extra time to cover the integer boundaries. A latency of five-second is feasible by using the zero-latency whitening and advanced computing hardware. The latency could be reduced to sub-second in an ideal scenario when all pipeline components reach the best latency solutions.

\begin{figure}
\includegraphics[width=\columnwidth]{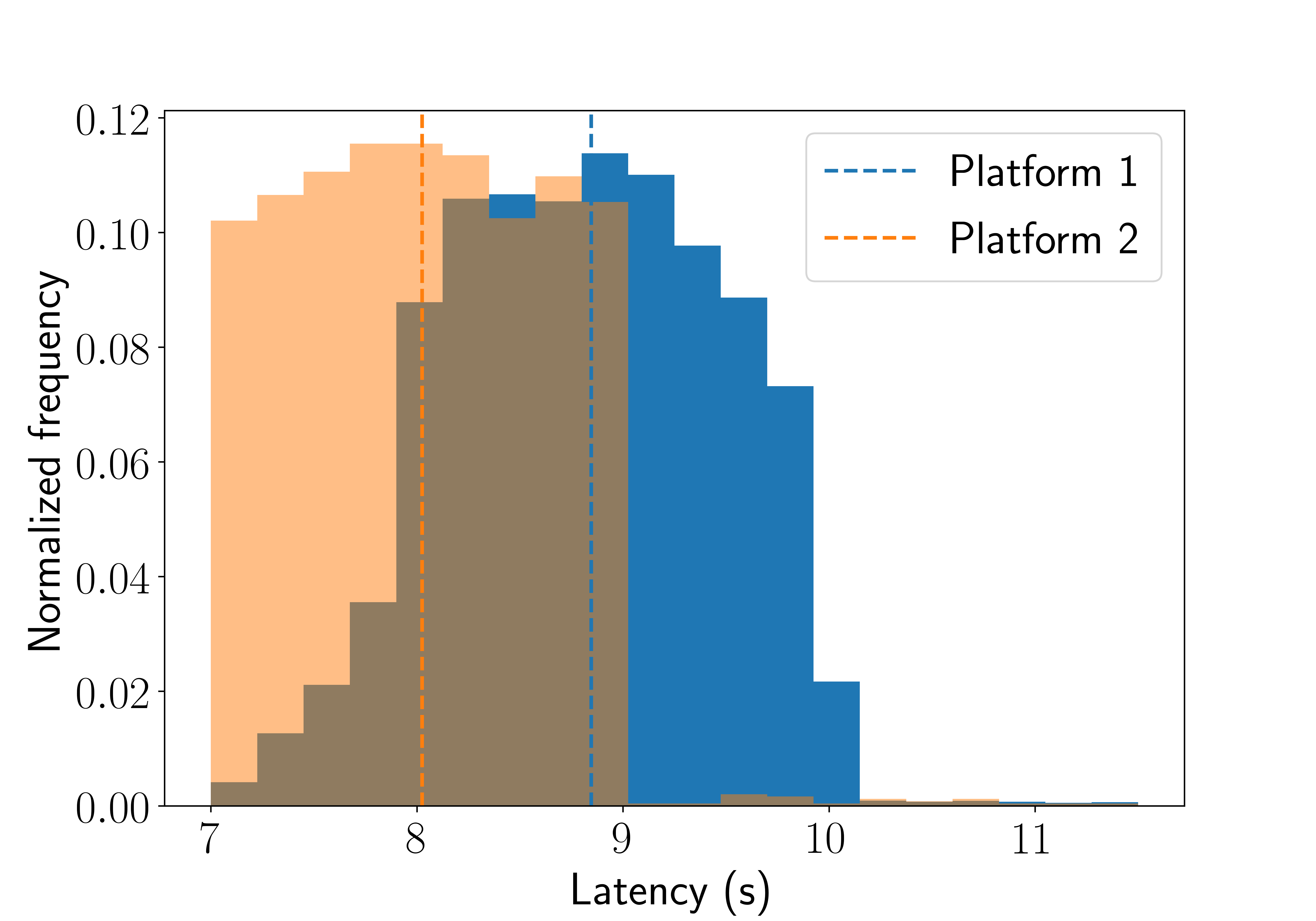}
\caption{Latency of the O3 SPIIR pipeline on two hardware platforms. Platform 1: one Quad-Core AMD Opteron(tm) 2376 CPU and one Nvidia GTX 1050Ti GPU. Platform 2: one Intel(R) Xeon(R) E5-2630 v4 @ 2.20GHz CPU and one Nvidia Titan V100 GPU. Dashed lines are median latencies. Platform 1 was used for the SPIIR pipeline for O3 online search.}
\label{fig:latency}
\end{figure}

To measure the latency of the pipeline in reality, we ran the pipeline over one-day stream replay of the O2 online data. We compared the pipeline latency on two computing platforms. As the CBC search is computational intensive and most of the computation has been carried out on GPUs, we listed the GPUs with their host CPUs for each platform. Platform 1 is a computing node loaded with relatively old hardware. Platform 2 is a computing node with state of art hardware. Just by using the latest computing hardware, the latency has been improved by nearly one second as shown by Fig.~\ref{fig:latency}. At the time of writing only the PyCBCLive pipeline published its O3 latency which is $>10$ seconds~\cite{DalCanton:2020vpm}. The median latency of the SPIIR pipeline is less than 9 seconds.

\begin{table}[]
    \centering
    \begin{tabular}{|p{0.3\columnwidth}|p{0.33\columnwidth}|p{0.33\columnwidth}|}
    \hline
      Component & Current latency (s) & Future latency (s) \\
      \hline
      Data package  & $1^{(1)}$ & ${N_{\textsc{\tiny PACK}}/N_\textsc{\tiny rate}}^{(1)}$  \\
      Downsampling  & ${N_{\textsc{\tiny FIR,D}}/N_{\textsc{\tiny rate}}}^{(1)}$ & ${N_{\textsc{\tiny FIR,D}}/N_{\textsc{\tiny rate}}}^{(1)}$ \\
      Whitening & $2^{(1)}$ & 0 \\
      Gating & $0.25^{(1)}$ & $0.25^{(1)}$ \\
      SPIIR filtering & $< 1^{(2)}$ & $< 1^{(2)}$ \\
      Coherent search & $<1^{(2)} + {N_{j}/N_\textsc{\tiny rate}}^{(1)}$ & $< 1^{(2)}+ {N_{j}/N_\textsc{\tiny rate}}^{(1)}$\\
      Trigger generation & $<1^{(2)} + 0.5^{(1)}$ & $< 1^{(2)}$ \\
      Data irregular slicing & $<1^{(1)}$ & $\sim 0$ \\
      Computing time from components other than afore mentioned & $<1^{(2)}$ & $\sim 0$ \\
      \hline
      Overall & $ <9$ & $<5$ \\
    \hline
    \end{tabular}
    \caption{Breakdown of the pipeline latency. ${}^{(1)}$ marks the intrinsic delay associated with data collection and ${}^{(2)}$ marks the latency associated with computing time. $N_{\text{rate}}$ is the data sample rate. $N_{\textsc{\tiny PACK}}$ is the number of samples for each data packet. $N_{\textsc{\tiny FIR,D}}$ is the filter length for downsampling. $N_j$ is the number of data samples used for $\xi^2$ calculation in the coherent search.}
    \label{tab:latency}
\end{table}

\subsection{O3 public alerts}
During O3, 80 public alerts were issued from candidates reported by the five online pipelines~\footnote{\url{https://gracedb.ligo.org/superevents/public/O3/}}. The rate of glitches predominantly at frequencies below 100 Hz has increased significantly during O3~\cite{gwtc2}. This resulted in 24 immediate retractions of  of submitted public alerts based on online investigation of noise association with the candidates. Two retractions were from the SPIIR pipeline on the same day of Jan. 6, 2020 due to high-amplitude scattered noise in L1 detector on that day. It is expected the offline searches using cleaned data will find some alerts to be insignificant and in the meantime rediscover some non-alerted online triggers with improved significance. Of the 56 online alerts, the SPIIR pipeline registered triggers that are associated with 38 alerts and published in LIGO-Virgo GCN notices~\footnote{\url{https://gcn.gsfc.nasa.gov/lvc\_events.html}}. 36 of them were reported with other pipelines and one alert was reported only by the SPIIR pipeline (S190910d) which is proved no longer significant by the offline deep searches~\cite{gwtc2}. 

%Of the 19 non-SPIIR reported alerts, 15 alerts were from triggers that presented high SNRs in one detector. One alert was reported only by the cWB burst pipeline indicating a higher-mass BBH system than the current coverage of CBC pipelines. Another two SPIIR triggers were produced but not uploaded to GraceDB due to technical issues. The GW190814 event was detected by the online SPIIR pipeline with consistent SNRs but was not uploaded due to pipeline veto on the Hanford detector data at that time. 

% As of the time of writing, the pipeline uploaded online triggers for the exceptional GW events ~\cite{gw190814}, ~\cite{gw190521} and ~\cite{gw190425}. The GW190814 event was detected by the SPIIR pipeline with consistent SNRs but was vetoed by the pipeline checks on the Hanford detector at that time.
 
\section{Conclusion and discussion}
\label{sec:conclusion}
This paper presents a low-latency pipeline used in the third LIGO-Virgo observation (O3) run, the SPIIR pipeline. It uses the efficient time-domain GPU-accelerated SPIIR method to perform time-domain matched filtering covering templates from binary neutron stars to intermediate binary black holes. It is the first low-latency CBC pipeline that employs the coherent search method with the help of GPU acceleration.  The median latency of the pipeline for O3 is less than nine seconds which is the fastest among online pipelines. It has the potential to be reduced to be below five seconds. %It provides sky localization for the candidates using a computational efficient Bayesian method. 

For the next coming run of O4, it will use data from new detector KAGRA and explore new algorithms to utilize the null-SNR information. Though the pipeline is using the novel coherent statistic for multi-detector detections, the single-detector significance estimation has been implemented only for vetoing but can be adapted to single detector triggering. This can help increase the number of detections up to 30\%~cite{gwtc2}.The early-warning configuration of the pipeline has been tested recently~\cite{Magee2021}. Given that the detector sensitivity is expected to improve significantly in this decade, early-warning detections have the potential to lead to major breakthroughs in the near future.

Additional information of the pipeline document and code can be found in \href{https://git.ligo.org/lscsoft/spiir/}{https://git.ligo.org/lscsoft/spiir/}.

\begin{acknowledgments}

This work was funded by the Australian Research Council (ARC) Centre of Excellence for Gravitational Wave Discovery OzGrav under grant
CE170100004. KK is partially supported by the National Research Foundation of Korea (NRF) grant funded by the Korea government (MSIT) (NRF-2020R1C1C1005863). TGFL was partially supported by grants from the Research Grants Council of the Hong Kong (Project No. CUHK14306419 and CUHK14306218), Research Committee of the Chinese University of Hong Kong and the Croucher Foundation of Hong Kong. A.Sengupta thanks the Department of Science and Technology for their ICPS cluster grant DST/ICPS/Cluster/Data\_Science/2018/General/T-150. We wish to acknowledge Tom Almeida, Andrew Munt, Zhaohong Peng, Han-Shiang Kuo, Fengli Lin, Guo Chin Liu for helpful discussions on the improvement of the work. 

This work used the computer resources of the LIGO CIT (Caltech) computer cluster and OzStar computer cluster at Swinburne University of Technology. LIGO CIT cluster is funded by National Science Foundation Grants PHY-0757058 and PHY-0823459. The OzSTAR program receives funding in part from the Astronomy National Collaborative Research Infrastructure Strategy (NCRIS) allocation provided by the Australian Government. We wish to thank Stuart Anderson, Jarrod Hurley for the great help to use the clusters. This research used data obtained from the Gravitational Wave Open Science Center (https://www.gw-openscience.org), a service of LIGO Laboratory, the LIGO Scientific Collaboration and the Virgo Collaboration. LIGO is funded by the U.S. National Science Foundation. Virgo is funded by the French Centre National de Recherche Scientifique (CNRS), the Italian Istituto Nazionale della Fisica Nucleare (INFN) and the Dutch Nikhef, with contributions by Polish and Hungarian institutes. This research used the injection sets generated by the rates and population group of the LIGO Scientific Collaboration. We acknowledge the GstLAL Team for the GstLAL library for several modules used in this work.

\end{acknowledgments}

\section{Appendixes}

\subsection{Network log likelihood ratio}
\label{appx:llr}
A GW signal is a function of two sets of parameters, the intrinsic parameters ($\bs{\Theta}$) including the masses and spins of the components, and extrinsic parameters including the sky location (right ascension $\alpha$ and declination $\delta$), the coalescence time $t_c$, the luminosity distance $l$, the polarization angle $\psi$, the inclination angle $\iota$, and the GW coalescence phase $\phi_c$.  Previous work have shown that four extrinsic waveform parameters ($l$, $\psi$, $\iota$, and $\phi_c$) can be analytically maximized for the network log likelihood ratio (LLR) statistic~\cite{pai01,bose2011,harry2011,macleod2016} which leaves a reduced set of parameters for the network LLR representation. Here we simplify the reduced LLR using the SVD technique and show that the SNRs from matched filtering can be directly used to construct the this network LLR statistic which is referred to the coherent SNR throughout the paper. % This reduced LLR is the coherent network SNR statistic used in the pipeline.

We first show the GW signal expression with an interferometric detector $I$:
\begin{equation}
\label{eq:strain_sngl}
 h_I(t) = F^{+}_I(t)h^+(t) + F^{\times}_I(t)h^{\times}(t),
\end{equation}
where $h^{+,\times}$ are the two polarizations and $F^{+,\times}$ are beam-pattern functions describing the responses of a detector to the two polarizations. 

By rearranging the extrinsic parameters, the signal can then be expressed with the direction-induced modulations $G^{+,\times}$ which is dependent on the source sky location $\alpha$ and $\delta$, and the detector location and orientation $\bs{s}(t)$; the ${a}_{jk}$ matrix pertaining to the source luminosity distance $l$, the polarization angle $\psi$, the inclination angle $\iota$, and the GW coalescence phase $\phi_c$; and the $h_c$ and $h_s$ waveforms which only depend on the intrinsic parameters and the coalescence time.

\begin{widetext}
\begin{equation}
%\begin{split}
    h_I(t;\bs{\Theta},\alpha,\delta, t_c,l, \psi, \iota,\phi_c, \bs{s}) =
    \left(G^+_I \left(\alpha, \delta, \bs{s}(t) \right)\Hquad G^{\times}_I \left(\alpha, \delta, \bs{s}(t)\right) \right) 
\left (
\begin{array}{cc}
a_{11} & a_{12} \\
a_{21} & a_{22}
\end{array}
\right )
\left (
\begin{array}{c}
h_c(t;\bs{\Theta},t_c) \\
h_s(t;\bs{\Theta},t_c)
\end{array}
\right ). \label{eq:hc}
%\end{split}
\end{equation}
\end{widetext}

The $G^{+,\times}$ expressions can be found in Eq.12 and Eq.13 of \cite{jaranowski98} ($G^{+}$ equivalent to $a(t)$ and $G^{\times}$ equivalent to $b(t)$ in cited equations) or Eq.1.53 and Eq.1.54 in~\cite{chichi_thesis}. They are related to the beam-pattern functions by the polarization angle $\psi$:
\begin{equation*}
\left ( \begin{array}{c}
F^{+}_I(t) \\
F^{\times}_I(t)
\end{array} \right) = 
\left ( \begin{array}{cc}
\cos 2\psi & \sin 2\psi \\
-\sin 2\psi & \cos 2\psi \\
\end{array} \right ) 
\left( \begin{array}{c} 
G^{+}_I(t) \\
G^{\times}_I(t) \\
\end{array} \right).
\label{ftwo}
\end{equation*}
The $a_{jk}$ matrix can be expressed as:
\begin{widetext}
\begin{equation}
\left (
\begin{array}{cc}
a_{11} & a_{12} \\
a_{21} & a_{22}
\end{array}
\right ) = 
\frac{1}{l}
\left ( \begin{array}{cc}
\cos 2\psi & \sin 2\psi \\
-\sin 2\psi & \cos 2\psi
\end{array} \right )
\left ( \begin{array}{cc}
\frac{1}{2}(1+\cos^2 \iota) & 0 \\
0 & \cos \iota
\end{array} \right )
\left ( \begin{array}{cc}
\cos \phi_c & \sin \phi_c \\
-\sin \phi_c & \cos \phi_c
\end{array} \right ).
\end{equation}
\end{widetext}
The solutions to $\{l,\psi,\iota,\phi_c\}$ from $a_{jk}$ can be found in ~\cite{bose2011}. $h_c$ and $h_s$ are in quadrature and the strength of each at a distance of 1 Mpc seen from a detector is defined by $\sigma_I$:
\begin{equation}
\begin{array}{c}
 (h_c \mid h_s) = 0, \\
 \sigma^2_I \equiv \frac{1}{1\text{Mpc}}(h_c \mid h_c) = \frac{1}{1\text{Mpc}} ( h_s \mid h_s ).\\
\end{array}
\label{eq:h_prop}
\end{equation}
The operator $(\cdot \mid \cdot)$ is defined as:
\begin{equation}
  (a\mid b)= 4 \text{Re} \int_{0}^{\infty} \frac{\tilde{a}(f)\tilde{b}^*(f)}{S_{n_I}(f)}\mathrm{d}f, 
  \label{eq:inner_product}
\end{equation} where $S_{n_I}(f)$ is the noise PSD in this detector.

% so that $(h_T|h_T) = 2$. $\text{Mpc}$ is a distance unit, mega Parsec. 

The network LLR is the sum of single LLRs assuming the noises in individual detectors are independent:
\begin{eqnarray}
   \ln{\mc{L}_{\text{NW}}} &=& \sum_{I} (d_I | h_I) - \frac{1}{2} (h_I | h_I),\nonumber \\
   &= & (\bs{d}^T|\bs{h}) -\frac{1}{2}(\bs{h}^T|\bs{h}), \label{eq:llr_def}
\end{eqnarray}
where the subscript $\text{NW}$ stands for the network, $\bs{d} = (d_1,...,d_{N_d})^T$ with $N_d$ being the number of detectors. $\bs{h} = (h_1,...,h_{N_d})^T$ depends on the detector location. The operator on the matrix is defined as:
\begin{equation}
\left( \bs{D} \mid \bs{B} \right) = \sum_{p=1}^n \left (D_{jp} \mid B_{pk} \right)_{p}.
\end{equation}

%From $h_c$ and $h_s$ one construct the CBC search template:
%\begin{equation}
% h_T(t;\bs{\Theta},t_c) = \frac{1}{\sigma} \left( h_c\left(t;\bs{\Theta},t_c \right) + i h_s \left(t;\bs{\Theta},t_c\right)\right) 
%\end{equation}
%where $i$ is the complex symbol and $\sigma$ is defined as:

%The extrinsic parameters, {$\alpha,\delta,\iota,l,\psi,\phi_c$} can be expressed with two parameters, the effective distance $l_\mathrm{eff}$ and a phase $\phi_0$~\cite{bruce05}~\cite{brue05}. The expression of the signal $h$ can be then simplified as:
%\begin{equation}
%h(t;\bs{\Theta},l_{\text{eff}},\phi_0) = \frac{1}{l_{\text{eff}}} h_c(t;\bs{\Theta},\phi_0),
%\end{equation}
%and $h_c$ is expanded with amplitude and phase as:
%\begin{equation}
%h_c(t;\bs{Theta},\phi_0) = \text{Re} A(t_c-t) e^{i \phi(t_c-t)}
%\end{equation}
%\text{Re}\Hquad A(t_c-t) e^{i \phi(t_c-t)}

We group the $a_{ij}$ into two entities $\bs{A}_c=(a_{11}, a_{21})^T$ and $\bs{A}_s=(a_{12}, a_{22})^T$ for expression convenience. $\bs{G}$ represents the modulation for each detector. The network LLR can be written as:
\begin{widetext}
\begin{equation}
\ln \mc{L}_{\text{NW}} = \left ({\bs{d}}^T\mid {\bs{G}}\bs{A}_c h_c + {\bs{G}}\bs{A}_s h_s \right) 
 -\frac{1}{2} \left ( \bs{A}_c^T {\bs{G}}^T h_c + \bs{A}_s^T {\bs{G}}^T h_s \mid {\bs{G}} \bs{A}_c h_c + {\bs{G}} \bs{A}_s h_s \right ). \label{eq:llr_extend}
\end{equation}
\end{widetext}

Maximizing LLR over $a_{jk}$ is equivalent to maximization over $\bs{A}_c$ and $\bs{A}_s$ respectively. The solution is then:
\begin{eqnarray}
\bs{A}_{x}\bigg|_{x=\{c,s\}}  &=& \left ( {\bs{G}}^T h_x \mid {\bs{G}} h_x \right)^{-1} {\bs{G}}^T ({\bs{d}^T} \mid h_x), \nonumber \\
&=& \frac{1}{1\mathrm{Mpc}}  (\bs{G}^T_{\sigma} \bs{G}_{\sigma})^{-1} \bs{G}^T_{\sigma} \left( \overbar{\bs{H}}_x \mid {\bs{d}^T}  \right ), \label{eq:Ax}
\end{eqnarray}
where
\begin{equation}
\overbar{\bs{H}}_x\bigg|_{x=\{c,s\}} = \left (
\begin{array}{cccc}
h_{x}/\sigma_{1} & 0 & \ldots & 0\\
0 & h_{x}/\sigma_{2} & \ldots & 0\\
\vdots & \vdots & \ddots & \vdots \\
0 & 0 & \ldots & h_{x}/\sigma_{N_d}\\
\end{array} \right ).
\end{equation}
$\bs{G}_{\sigma}$ is the noise-weighted modulation and its SVD has the form:
\begin{equation}
{\bs{G}}_{\sigma} = \left (
\begin{array}{cc}
G_{1}^{+}\sigma_{1} & G_{1}^{\times}\sigma_{1} \\
G_{2}^{+}\sigma_{2} & G_{2}^{\times}\sigma_{2} \\
\vdots & \vdots \\
G_{N_d}^{+}\sigma_{N_d} & G_{N_d}^{\times}\sigma_{N_d}
\end{array} \right ) =
\bs{U} \bs{\Lambda} \bs{V}^T, \bs{\Lambda} = \left (
\begin{array}{cc}
\lambda_1 & 0 \\
0 & \lambda_2 \\
\vdots & \vdots \\
0 & 0
\end{array}
\right ).
\label{eq:svd}
\end{equation}

The SVD decompose $\bs{G}_{\sigma}$ into a $N_d\times N_d$ unitary matrix $\bs{U}$, a $N_d\times 2$ pseudo-diagonal matrix $\bs{\Lambda}$ with decreasing positive singular values, and the transpose of a $2\times 2$ unitary matrix $\bs{V}$. This form can be used obtain the Moone-Penrose pseudo inverse $( {\bs{G}}^T h_x \mid {\bs{G}} h_x )^{-1}$ and simplify Eq.~\ref{eq:Ax}.

Substituting the solutions of $\bs{A}$ into the network LLR, the maximized LLR can then be expressed as:
\begin{eqnarray}
\underset{\max \{ \bs{a}_{jk} \} }{\ln \mc{L}_{\text{NW}}} & = & \frac{1}{1\mathrm{Mpc}} \sum_{x=\{c,s\}}\left (\overbar{\bs{H}}_x \mid {\bs{d}} \right )^T \bs{U}\bs{I}\bs{U}^{T} \left (\overbar{\bs{H}}_x \mid {\bs{d}} \right ) \nonumber \\
  & = &\frac{1}{1\mathrm{Mpc}} \parallel \bs{I}\bs{U}^T \left (\overbar{\bs{H}}_c + i\overbar{\bs{H}}_s \mid {\bs{d}} \right ) \parallel^2 \\
  & = & \frac{1}{1\mathrm{Mpc}}  \left\lVert \bs{I}\bs{U}^T \left (
\begin{array}{cccc}
z_1 & 0 & \ldots & 0\\
0 & z_2 & \ldots & 0\\
\vdots & \vdots & \ddots & \vdots \\
0 & 0 & \ldots & z_{N_d}\\
\end{array}
\right )
\right\rVert^2,\label{eq:llr_form}
\end{eqnarray}
where $\parallel \cdot \parallel$ is the Euclidean norm, $\bs{I} = \text{diag}\{1,1,0,..0\}$, and $z$ is the individual SNR from each detector. The maximization procedure can be thought of as a projection of all the signal components in the $N_d$ streams onto the signal plane spanned by the two vectors from $\bs{U}$, with the noise contributions reduced from $N_d$ Gaussian streams to two Gaussian streams. If the noise is Gaussian in each detector, then this statistic will obey a non-central $\chi^2$ distribution with a 4 degrees of freedom.

For a detector network with more than two detectors, the null stream or the null statistic can then be expressed as:
\begin{equation}
\ln \mc{L}_{\text{NULL}} = \frac{1}{1\mathrm{Mpc}}  \left\lVert \bs{I}^{\dagger} \bs{U}^T \left (
\begin{array}{cccc}
z_1 & 0 & \ldots & 0\\
0 & z_2 & \ldots & 0\\
\vdots & \vdots & \ddots & \vdots \\
0 & 0 & \ldots & z_{N_d}\\
\end{array}
\right ) \right\rVert^2,
\end{equation}
where $\bs{I}^{\dagger} = diag\{0,0,1,\ldots,1\}$. When noise is Gaussian in each detector this statistic follows a central $\chi^2$ distribution with $(N_d\times2 -4)$ degrees of freedom. %This information can then be used to veto glitches from environmental sources as described in~\cite{harry2015coh}.

\bibliographystyle{unsrt_sl}
% BIB
\bibliography{ref}% Produces the bibliography via BibTeX.

\end{document}